\documentclass[preprints,article,accept,moreauthors,pdftex]{mdpi}

\firstpage{1} 
\pubvolume{1}
\issuenum{1}
\articlenumber{0}
\pubyear{2022}
\copyrightyear{2022}
\externaleditor{Academic Editor: Marc S. Seigar} 
\datereceived{2 May 2022} 
\dateaccepted{26 May 2022} 
\datepublished{} 
\hreflink{https://doi.org/} 

\usepackage{graphicx}	
\usepackage{amsmath}	
\usepackage{amssymb}	

\newcommand{\A}{\text{\normalfont\AA}}

\def \hbeta{H$\beta$}

\def \halpha{H$\alpha$}
\def \Msol{{M}_{\odot}}

\def \Lsol{{\rm L}_{\odot}}

\def \logm{\log(M/\Msol)}
\def \lya{Ly$\alpha$}

\def \h2{{\rm H_{2}}}

\def \cii{C$^+$}

\def \oii{[O{\scriptsize ~II}]}

\def \nii{[N{\tiny\,II}]}
\def \hii{H{\tiny\,II}}

\def \LIR{L_{{\rm IR}}}
\def \LFIR{L_{{\rm FIR}}}
\def \LUV{L_{{\rm UV}}}

\def \LCII{L_{{\rm C^{+}}}}

\def \IRXB{IRX$-\beta$}
\def \IRXM{IRX$-$M$_*$}
\def \dn4000{D_{{\rm n}}(4000) }

\def \arcsec{^{\prime\prime}}

\newcommand\araa{Annu. Rev. Astron. Astrophys.}
\newcommand\apj{ Astrophys. J.}
\newcommand\apjl{ Astrophys. J.}
\newcommand\apjs{ Astrophys. J. Suppl. Ser.}

\newcommand\aap{Astron. Astrophys.}

\newcommand\mnras{Mon. Not. R. Astron. Soc.}

\newcommand\pasj{PASJ}

\newcommand\nat{Nature}

\def \gt{>}
\def \lt{<}

\Title{ALPINE: A Large Survey to Understand Teenage Galaxies}

\TitleCitation{ALPINE: A Large Survey to Understand Teenage Galaxies}

\Author{Andreas L. 
\textls[-15]{Faisst $^{1,*}$\orcidA{}, Lin Yan $^{2}$\orcidB{}, Matthieu B\'ethermin $^{3}$\orcidD{}, Paolo Cassata $^{4,5}$\orcidE{}, Miroslava  Dessauges-Zavadsky $^{6}$\orcidI{}, Yoshinobu Fudamoto $^{6,7,8}$\orcidC{}, Michele Ginolfi $^{9}$\orcidL{}, Carlotta Gruppioni $^{10}$\orcidH{}, Gareth Jones $^{11,12}$\orcidJ{}, Yana  Khusanova $^{13}$\orcidO{}, Olivier LeF\`evre $^{3}$\orcidN{}, Francesca  Pozzi $^{10,14}$\orcidM{}, Michael Romano $^{15}$\orcidF{}, Daniel Schaerer $^{6}$\orcidP{}, John Silverman $^{16,17}$\orcidK{}, and Brittany Vanderhoof $^{18}$\orcidG{}}}

\AuthorNames{A. Faisst, L. Yan, M. Bethermin, P. Cassata, M. Dessauges-Zavadsky, Y. Fudamoto, M. Ginolfi, C. Gruppioni, G. Jones, Y. Khusanova, O. LeFevre, F. Pozzi, M. Romano, D. Schaerer, J. Silverman, and B. Vanderhoof}

\AuthorCitation{Faisst, A.; Yan, L.; B\'ethermin, M.; Cassata, P.; Dessauges-Zavadsky, M.; Fudamoto, Y.; Ginolfi, M.; Gruppioni, C.; Jones, G.; Khusanova, Y.; et~al.}

\address{%
$^{1}$ \quad IPAC, M/C 314-6, California Institute of Technology, 1200 East California Boulevard, Pasadena,~CA~91125,~USA\\
$^{2}$ \quad The Caltech Optical Observatories, California Institute of Technology, Pasadena, CA 91125, USA; lyan@caltech.edu\\
$^{3}$ \quad Laboratoire d'Astrophysique de Marseille, 38 rue Fr\'ed\'eric Joliot-Curie, F-13013 Marseille, France; matthieu.bethermin@lam.fr\\
$^{4}$ \quad Dipartimento di Fisica e Astronomia, Universit\`a di Padova, Vicolo dell'Osservatorio 3, I-35122 Padova, Italy; paolo.cassata@unipd.it\\
$^{5}$ \quad INAF---Osservatorio Astronomico di Padova, Vicolo dell'Osservatorio 5, I-35122 Padova, Italy \\
$^{6}$ \quad Observatoire de Genev\`e, Universit\'e de Geneve, 51 Ch. des Maillettes, CH1290 Versoix, Switzerland; miroslava.dessauges@unige.ch (M.D.-Z.); y.fudamoto@aoni.waseda.jp (Y.F.)\\
$^{7}$ \quad Research Institute for Science and Engineering, Waseda University, 3-4-1 Okubo, Shinjuku~169-8555,~Tokyo
,~Japan \\
$^{8}$ \quad National Astronomical Observatory of Japan, 2-21-1, Osawa, Mitaka
,~Tokyo,~Japan \\
$^{9}$ \quad European Southern Observatory, Karl-Schwarzschild-Strasse 2, 85748 Garching, Germany; michele.ginolfi@eso.org\\
$^{10}$ \quad \hspace{-0.3em}Osservatorio di Astrofisica e Scienza Dello Spazio (INAF-OAS), Via P. Gobetti 93/3, I-40129 Bologna, Italy; carlotta.gruppioni@inaf.it (C.G.); f.pozzi@unibo.it (F.P.)\\
$^{11}$ \quad \hspace{-0.3em}Kavli Institute for Cosmology, University of Cambridge, Madingley Road, Cambridge CB3 0HA, UK; gj283@cam.ac.uk\\
$^{12}$ \quad \hspace{-0.3em}Cavendish Laboratory Astrophysics Group, University of Cambridge, 19 J. J. Thomson Ave., Cambridge~CB3~0HE,~UK \\
$^{13}$ \quad \hspace{-0.3em}Max-Planck-Institut für Astronomie, Königstuhl 17, D-69117 Heidelberg, Germany; khusanova@mpia.de\\
$^{14}$ \quad \hspace{-0.3em}Dipartimento di Fisica e Astronomia, Universit\` degli Studi di Bologna, Via P. Gobetti 93/2, I-40129~Bologna,~Italy \\
$^{15}$ \quad \hspace{-0.3em}National Centre for Nuclear Research, Ul. Pasteura 7, 02-093 Warsaw, Poland; michael.romano@ncbj.gov.pl\\
$^{16}$ \quad \hspace{-0.3em}Kavli Institute for the Physics and Mathematics of the Universe (Kavli IPMU, WPI)
, The University of Tokyo, Kashiwa 277-8583, Japan; silverman@ipmu.jp\\
$^{17}$ \quad \hspace{-0.3em}Department of Astronomy, School of Science, The University of Tokyo, 7-3-1 Hongo, Bunkyo~113-0033,~Tokyo,~Japan \\
$^{18}$ \quad \hspace{-0.3em}School of Physics and Astronomy, Rochester Institute of Technology, Rochester, NY 14623, USA; bxv3026@rit.edu\\
}

\corres{Correspondence: afaisst@caltech.edu}

\abstract{A multiwavelength study of galaxies is important to understand their formation and evolution. Only in the recent past, thanks to the {\it Atacama Large (Sub) Millimeter Array} (ALMA), were we able to study the far-infrared (IR) properties of galaxies at high redshifts. In this article, we summarize recent research highlights and their significance to our understanding of early galaxy evolution from the \textit{ALPINE} survey, a large program with ALMA to observe the dust continuum and $158\,{\upmu \rm{m}}$ \cii~emission of normal star-forming galaxies at $z=$~4--6. Combined with ancillary data at UV through near-IR wavelengths, \textit{ALPINE} provides the currently largest multiwavelength sample of post-reionization galaxies and has advanced our understanding of \textit{(i)} the demographics of \cii~emission; \textit{(ii)} the relation of star formation and \cii~emission; \textit{(iii)} the gas content; \textit{(iv)} outflows and enrichment of the intergalactic medium; and \textit{(v)} the kinematics, emergence of disks, and merger rates in galaxies at $z>4$.  \textit{ALPINE} builds the basis for more detailed measurements with the next generation of telescopes, and places itself as an important post-reionization baseline sample to allow a continuous study of galaxies over $13$ billion years of cosmic time.
}


\begin{document}



\section{Introduction}

The Universe that we observe today is diverse, with a mix of star-forming and quiescent galaxies with different internal (stellar mass, metallicity) and external (structure, environment) properties. While most star-forming galaxies (often called ``spiral galaxies'' due to their spiral structure) form stars at low rates of a few solar masses per year, starbursts exhibiting multiples of these rates are also observed. Quiescent galaxies on the other hand (called ``spheroids'' due to their mostly symmetric and compact appearance with a smooth light profile) had their star formation stopped in the past by a variety of reasons \citep{Ciotti1997,Ilbert2010,Cheung2016}. At earlier cosmic times, however, this picture changes; the sighting of quiescent galaxies is rare, and star-forming galaxies (showing an irregular and lumpy structure) form stars at rates higher than present-day starbursts. Observing the evolution of galaxies across cosmic times starting from early formation epochs and deriving a physical model of the evolutionary processes that explain the diversity of galaxies is one of the main goals of the current efforts in the astrophysical community.

From the combination of imaging and spectroscopic observations at ultraviolet (UV), optical, and (far-) infrared (IR) wavelengths, a reasonably robust picture of how galaxies evolve across the past 9--10 Gyrs of cosmic time (corresponding to redshifts $z < 3$) has been established. Similar to the local universe, galaxies exhibit a tight relation between star-formation rate (SFR) and stellar mass (the star-forming main-sequence; \citep{rodighiero11,SPEAGLE14}) as found by robust total SFR measurements from far-IR observations. However, galaxies in the earlier universe are significantly more star-forming on average and the fraction of quiescent galaxies is decreasing strongly at redshifts higher than $\sim$$2$ \citep{Bell2004,Ilbert2010,MADAU14}. This is also the epoch when the cosmic SFR density reaches the peak over cosmic time \citep{MADAU14}. Along with changing SFRs, the environment in which stars are being formed evolves across cosmic time. Galaxies become more dust- and metal-enriched over time as their stellar masses are being built up by galaxy--galaxy mergers and the inflow of new gas to fuel star formation \citep{Hopkins2006,Ho2015}. The rate of galaxy--galaxy major mergers (with a mass ratio smaller than 1:4) as well as the gas fractions of galaxies are observed to increase towards earlier cosmic times \citep{Tacconi2020}. The overall increase in SFR as a function of redshift is found to be coupled with increasingly large gas reservoirs, while the vertical position on the main-sequence at a given stellar mass and redshift is dominated by changes in star-formation efficiency \citep{SCOVILLE17,SILVERMAN15}. Finally, high-resolution space-based observations of the light distribution of galaxies at early times show that they are significantly more irregular and dominated by clumps of star-forming gas and minor mergers \citep{Conselice2014,BOUWENS14,ELMEGREEN2021}. Spectroscopic observations of the \halpha~emission line of large samples of galaxies at $z$$\sim$$2$ have confirmed this picture by revealing a significant amount of dispersion dominated and merging galaxies (e.g., \citep{FORSTERSCHREIBER11}).

\begin{figure}[t]
\includegraphics[width=0.8\textwidth]{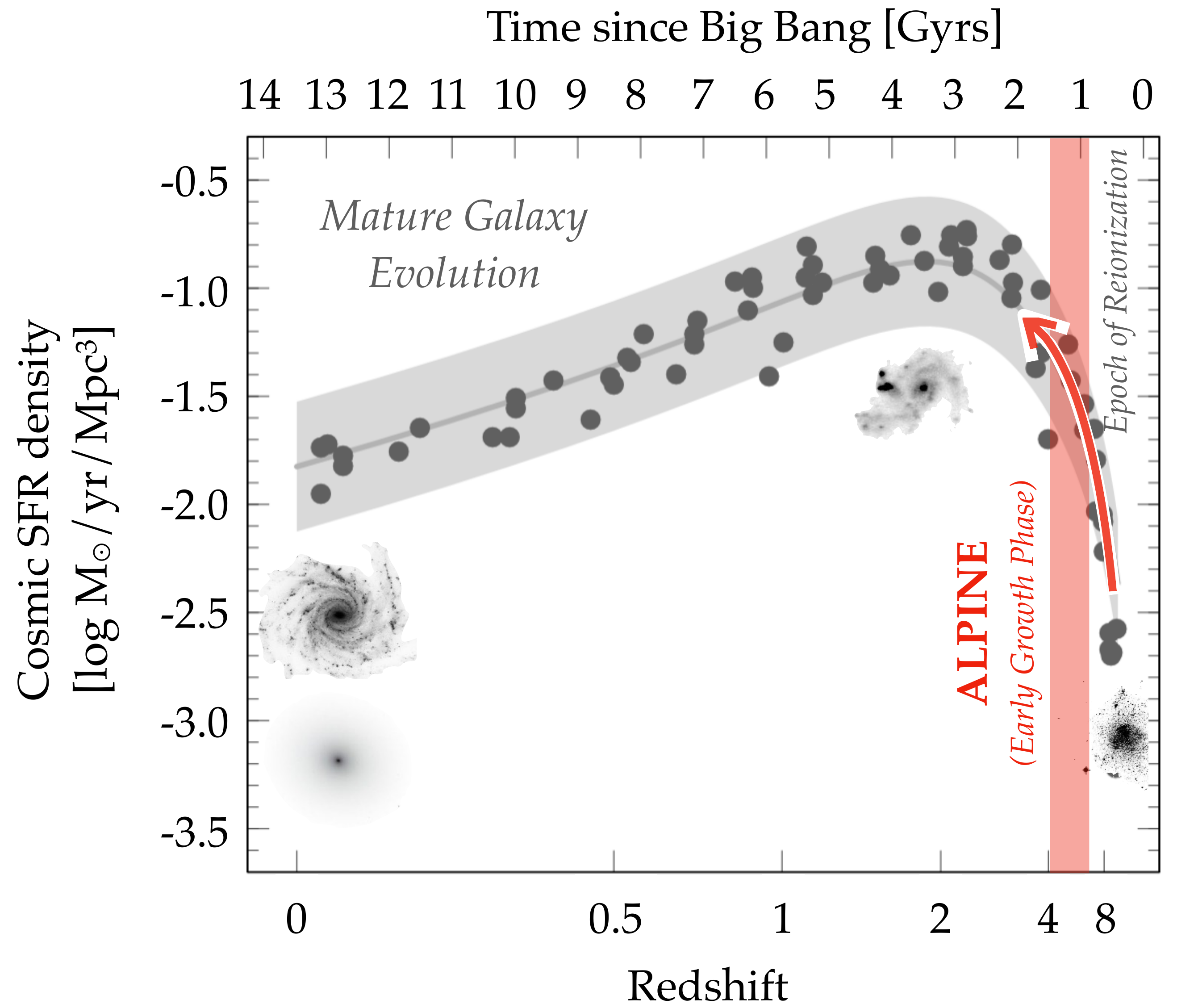}
\caption{Cosmic SFR density 
 as a function of cosmic time and redshift. The global SFR density increases sharply during primordial galaxy assembly and evolution to reach a peak at $z$$\sim$2--3 followed by a decline due to the emergence of quiescent galaxies during the mature galaxy evolution. The \textit{ALPINE} survey (presented in
 \citet{ALPINE_BETHERMIN19}, \citet{ALPINE_FAISST20}, and \citet{ALPINE_LEFEVRE20}) studies galaxy evolution during the Early Growth Phase at $z=$~4--6, an important connection between primordial and mature galaxy evolution, via a large multiwavelength sample. \textit{ALPINE} builds the basis for more detailed measurements with the next generation of telescopes, and places itself as an important post-reionization baseline sample to allow a continuous study of galaxies over $13$ billion years of cosmic time.
The grayscale images show the change in structure of the galaxies across cosmic time. Data points from \citet{MADAU14} and \citet{BOUWENS15}.
\label{fig:sfrd}}
\end{figure}

Large spectroscopic campaigns using 8--10 m-class telescopes over the past decade have extended the study of statistical samples of galaxies beyond the peak of cosmic SFR density towards the Epoch of Reionization at $z$$\sim$$6$. The largest of these programs use the VIMOS spectrograph on the \textit{Very Large Telescope} (VLT) \citep{LEFEVRE03,LEFEVRE15} and the DEIMOS spectrograph on the \textit{Keck} telescopes \citep{FABER03,HASINGER18}, and provide robust redshifts for hundreds of galaxies at $z=$~4--6. This redshift range, the ``Early Growth Phase'' of galaxy evolution, represents a transition phase between primordial galaxy formation in the Epoch of Reionization and mature galaxy evolution at the peak of cosmic SFR density, and hence forms the crucial link between early galaxies and modern galaxies (Figure~\ref{fig:sfrd}).
In addition to spectroscopy, 10,000~s of galaxies have been identified photometrically in this redshift range over many extragalactic fields observed with the \textit{Hubble} space telescope in multiple photometric bands (e.g.,~\citep{BOUWENS15}). Large programs using the \textit{Spitzer} space telescope, covering near-IR wavelengths, have imaged the rest-frame optical light (emitted by older stars), which is crucial for estimating stellar masses of galaxies at $z>4$ \citep{DAVIDZON17} and even allows the measurement of \halpha, \oii, and \hbeta~emission lines \citep{Shim2011,FAISST16a}.

While such large samples at these early cosmic times exist, they are mostly selected at rest-frame UV (and sometimes optical) wavelengths. This means that these samples are highly biased towards galaxies visible at these wavelengths, {i.e.}, not or only slightly attenuated by dust. As estimated in \citet{BOUWENS09}, at $z=$~4--6, more than $40\%$ of the UV light is reprocessed by dust and another $10\%$ is completely missed by UV studies. This significantly limits our understanding of this galaxy population in UV light and, for example, raises the question of how much dust-obscured star formation contributes to the growth of these galaxies. Furthermore, the observation of the spatial and kinematic structure of the colder gas (only observable at submillimeter wavelengths) provides new insights into the physics of star-formation and the conditions of the interstellar medium (ISM) in these galaxies, complementary to what can be done at rest-frame UV and optical~wavelengths.

The high sensitivity of the \textit{Atacama Large (Sub-) Millimeter Array} (ALMA) enables such observations at sub-mm wavelengths. In combination with data in the UV and optical, it enables us to study early galaxy evolution in more detail than ever before. While bright sub-mm galaxies, including (Ultra) Luminous Infrared Galaxies (ULIRGs), have been observed before with \textit{Herschel} \citep{Lutz2011,Oliver2012}, ALMA extends the analysis to \textit{normal} main-sequence galaxies at early cosmic times by targeting their far-IR dust continuum together with emission lines, such as singly ionized carbon (\cii, one of the most dominant cooling lines) at rest-frame $158\,{\upmu \rm{m}}$.

The importance of such a multiwavelength approach was first demonstrated on a sample of $10$ galaxies observed in \cii~at $z$$\sim$5 in \citet{CAPAK15} (see also \citep{RIECHERS14,BARISIC17,FAISST17b}). All galaxies reside on the \textit{Cosmic Evolution Survey} (COSMOS, \citet{SCOVILLE07}) field; thus, a wealth of data is available
from UV to optical wavelengths, including rest-frame UV spectroscopy from \textit{Keck}/DEIMOS and \textit{VLT}/VIMOS \citep{HASINGER18,LEFEVRE15}. The coverage of the field with \textit{Hubble}/ACS imaging in F814W ($i$-band) \citep{KOEKEMOER07} as well as partially in the near-IR from the CANDLES survey \citep{GROGIN11,KOEKEMOER11}, offers further high-resolution spatial information. With $10$ out of $10$~\cii~detections, the study suggested that \cii~is abundant in this galaxy population and a good tool to study these galaxies.
In summary, the study revealed large variations in the dust content (measured by sub-mm continuum) and \cii~emission in this small sample, and in particular highlighted the kinematics of molecular gas and ``UV-dark'' dust-obscured major mergers, which have been missed previously.
However, without a large sample with similar multiwavelength measurement, further conclusions on the galaxy population as a whole are limited.

The \textit{ALMA Large Program to Investigate \cii~at Early Times} (ALPINE; \mbox{\citet{ALPINE_LEFEVRE20}}, \citet{ALPINE_BETHERMIN19}, \citet{ALPINE_FAISST20}) mitigates these issues by providing similar far-IR dust continuum and \cii~measurements for a sample of $118$ main-sequence galaxies spectroscopically confirmed at $4.4 < z < 5.9$ in the COSMOS field and the \textit{Extended Chandra Deep Field South} (ECDFS, \citet{GIACCONI02}). The \textit{ALPINE} ALMA data (described in \citet{ALPINE_BETHERMIN19}) are taken by a $69\,{\rm hour}$ ALMA large program over cycles 5 and 6. All galaxies have a wealth of multiwavelength data and rest-UV spectroscopy (described in \citet{ALPINE_FAISST20}), which makes \textit{ALPINE} the first and largest multiwavelength survey of galaxies at these early cosmic times.
\textit{ALPINE} has significantly advanced our understanding of post-reionization galaxies during the \textit{Early Growth Phase} at $z=$~4--6 in several topics,~including 

\begin{itemize}
    \item contribution of obscured star formation to the cosmic SFR density;
    \item gas reservoirs and modes of star formation;
    \item enrichment of the circumgalactic medium (CGM) via outflows from galaxies;
    \item mixture of kinematic structure, emergence of rotators, and merger rates;
    \item abundance of ``UV-dark'' sources detected serendipitously at high redshifts.
\end{itemize}

In this article, we summarize the research highlights from \textit{ALPINE} and their significance to our understanding of early galaxy evolution. Specifically, in Section~\ref{sec:alpineintro}, we provide a brief introduction of \textit{ALPINE} including its multiwavelength data products. In Section~\ref{sec:results}, we present in detail the science results and we discuss their significance in Section~\ref{sec:discussion}. We conclude in Section~\ref{sec:end}.

Throughout this work, we assume a $\Lambda$CDM cosmology with $H_0 = 70\,{\rm km\,s^{-1}\,Mpc^{-1}}$, $\Omega_\Lambda = 0.70$, and $\Omega_{\rm m} = 0.30$. All magnitudes are given in the AB system \citep{OKE74} and stellar masses and SFRs are normalized to a \citet{CHABRIER03} initial mass function (IMF) unless noted~otherwise.

\section{\textit{ALPINE}---The Largest Post-Reionization Multiwavelength Survey}\label{sec:alpineintro}

The primary goal of \textit{ALPINE} is to study the gas and dust properties of galaxies shortly after the Epoch of Reionization at $z$$\sim$4--6 via the $150\,{\upmu \rm{m}}$ dust-continuum and \cii~line emission in combination with rest-UV to optical observations. This epoch of cosmic time is of particular interest as it connects primordial galaxy formation at $z>6$ with mature galaxy evolution around the peak of cosmic SFR density ($z$$\sim$2--3). In this section, we summarize the sample and observations as well as the most important science goals. For more details on these topics, we refer the reader to \citet{ALPINE_BETHERMIN19} (ALMA observations) and \citet{ALPINE_FAISST20} (sample selection and non-ALMA ancillary data).

\subsection{Sample and Observations}

\begin{figure}[t]
\includegraphics[width=1\textwidth]{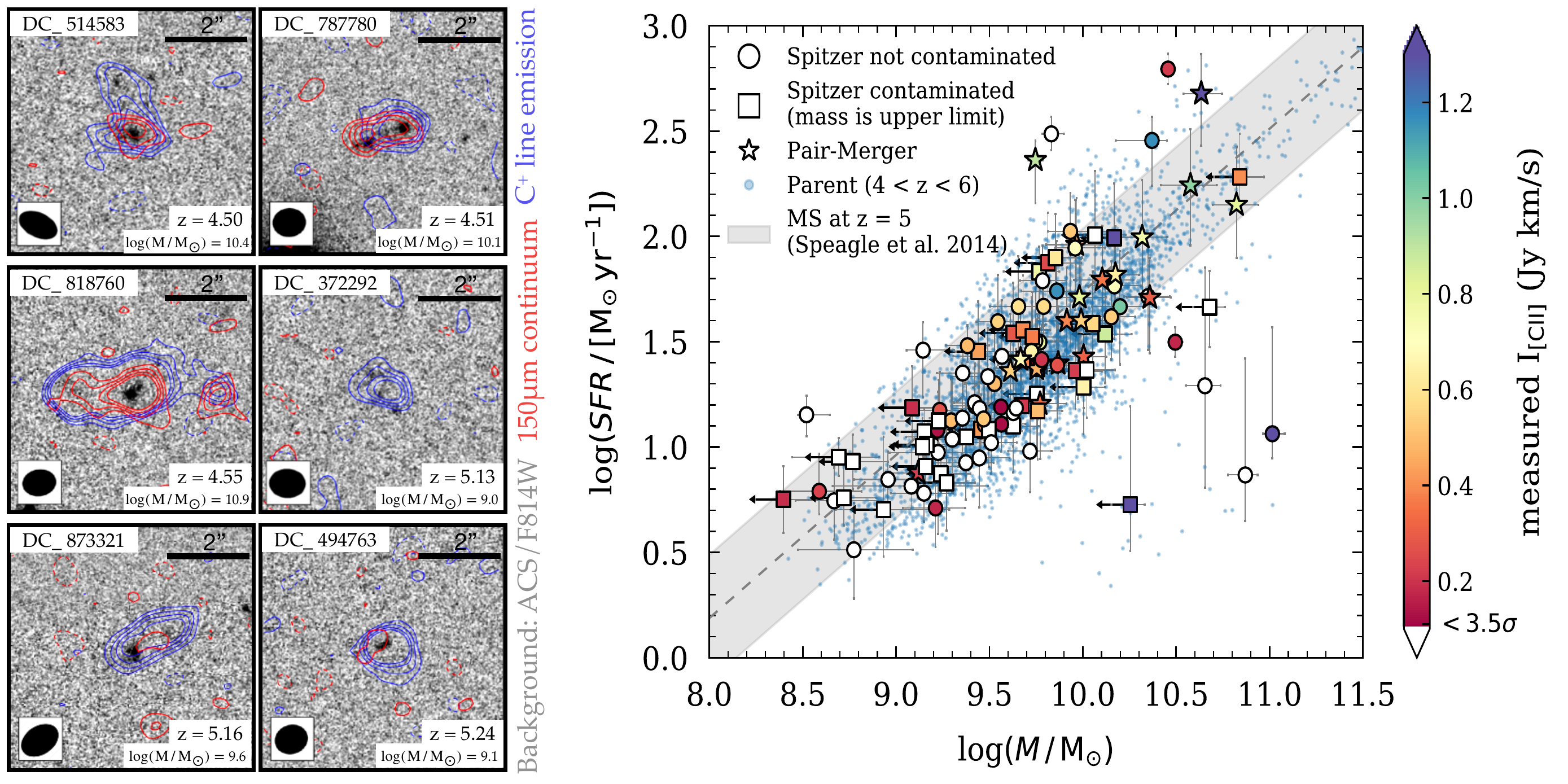}
\caption{\textit{ALPINE} observed $118$ galaxies on the star-forming main-sequence at redshifts \mbox{$z~=~$4.4--5.9}. The ALMA observations provide measurements of the $158\,{\upmu\rm{m}}$ \cii~line emission as well as the surrounding (rest-frame $\sim$$150\,\upmu$m) dust continuum. \textbf{Left Panels:} Hubble ACS/F814W cutouts of six \textit{ALPINE} galaxies at different redshifts with a range of \cii~and far-IR emission morphologies. Contours show \cii~emission (blue) and dust continuum (red), target IDs, stellar masses, and redshifts are also indicated. \textbf{Right Panel:} Distribution of the \textit{ALPINE} galaxies (large symbols) and their parent sample (blue dots) on the main-sequence (gray) at $z$$\sim$$5$. The color of the symbols indicate the observed \cii~line emission. Different symbols identify mergers (stars) as well as galaxies whose Spitzer photometry is contaminated by neighboring objects (squares); in which case, their stellar masses are likely overestimated. Adapted from Ref.
 \cite{ALPINE_FAISST20}.
\label{fig:mainsequence}}
\end{figure}

\textit{ALPINE} is a $69\,{\rm hour}$ ALMA large program ($\#$ 2017.1.00428.L) carried out in Cycles 5 and 6 in Band 7 (0.8--1.1 mm or 275--373 GHz). The program is a collaboration between several members of ALMA regional centers, including Europe, North America, Chile, and Japan. Of the $118$ galaxies with spectroscopic redshifts of 4.4--5.9, $105$ are located in the COSMOS field~\citep{SCOVILLE07} and $13$ in the ECDFS \citep{GIACCONI02}. The sample spans roughly two orders of magnitude in stellar mass and SFR on the star-forming main-sequence, hence including a representative sample of typical main-sequence galaxies at these cosmic times (Figure~\ref{fig:mainsequence}). The galaxies were selected in two redshift windows at $4.40 < z < 4.65$ (67 galaxies) and $5.05 < z < 5.90$ (51~galaxies) with median redshifts of $z$$\sim$$4.5$ and $z$$\sim$$5.5$, respectively. The windows were chosen to avoid frequencies of low atmospheric transmission and to optimize the observation efficiency. The spectroscopic redshifts were gathered from two large spectroscopic surveys carried out by \textit{Keck}/DEIMOS (\textit{DEIMOS 10k Survey}; \citep{HASINGER18}) and \textit{VLT}/VIMOS (\textit{VIMOS Ultra Deep Survey}, VUDS; \citep{LEFEVRE15}). These surveys combine redshifts of galaxies selected in a variety of ways, for example, by narrowband \lya~emission, Spitzer $4.5\,{\upmu \rm{m}}$ flux excess, Lyman Break color selection (LBGs), and photometric redshifts. This variety is inherited in the final \textit{ALPINE} sample, emphasizing the selection of representative galaxies at these redshifts. Furthermore, the redshifts were derived from \lya~emission as well as UV absorption lines, therefore minimizing selection biases towards young, dusty-poor \lya~emitters. All spectra have been consistently renormalized using the \textit{COSMOS2015} photometric catalog \citep{LAIGLE16} or CANDELS photometry, using broadband and narrowband data (see \mbox{\citet{ALPINE_FAISST20}} for a full description). The \textit{ALPINE} galaxies have an absolute rest-UV magnitude of $M_{\rm 1500} > -20.2\,{\rm mag}$, corresponding to an equivalent SFR of $10\,{\rm \Msol\,yr^{-1}}$. This cut was chosen to maximize the number of \cii~line detections according to the $\LCII$~vs. $M_{\rm 1500}$ relation found in \citet{CAPAK15} based on $10$ $z$$\sim$$5.5$ galaxies with similar~observations.

Most observations (102 galaxies) were carried out between May and August 2018 in Cycle 5 and the remaining ones were observed in January 2019 in Cycle 6. The \cii~lines of the targeted sources were covered by two contiguous spectral windows of $1.875\,{\rm GHz}$ each. The remaining two windows were used to observe the dust continuum around the \cii~emission line. The compact array configurations C43-1 and C43-2 were used to obtain a resolution $>0.7\arcsec$ to avoid over-resolving of the flux, hence decreasing the signal-to-noise (S/N). The final average synthesized beam full-width-at-half-maximum (FWHM), depending on frequency and exact array configuration, is $0.85\arcsec \times 1.13\arcsec$, corresponding to roughly 5--6 kpc at redshift $z=5$. The coarse spectral resolution is $\Delta v_{\rm channel}~=~31.250\,{\rm MHz}$, which allows the resolution of the \cii~emission line (typical FWHM of $235\,{\rm km\,s^{-1}}$).
The integration times per source vary from $10\,{\rm min}$ to $1\,{\rm h}\,20\,{\rm min}$, to achieve an average root mean square (rms) line sensitivity of $0.14\,{\rm Jy\,km\,s^{-1}}$ for \cii~emission\endnote{The \cii~emission was estimated from the $\rm L_{C^+}$ vs. $\rm M_{1500}$ relation from \citet{CAPAK15}.} (corresponding to $\LCII$$\sim$$0.4 \times 10^8\,\Lsol$) and rms continuum sensitivity of $50\,\upmu{\rm Jy\,beam^{-1}}$ for continuum. Out of the $118$ targeted galaxies, $23$ were detected in continuum at >$3.5\sigma$ ($\sim$$20\%$) and $75$ were detected in \cii~line emission at >$3.5\sigma$ ($\sim$$64\%$). For a full description of line and continuum measurements, we refer the reader to \citet{ALPINE_BETHERMIN19}. In addition, \textit{ALPINE} provides a blank-field survey of $\sim$$24\,{\rm arcmin^2}$, in which $57$ sources were detected serendipitously in continuum at >$5\sigma$ (described in \citet{ALPINE_BETHERMIN19}) and $14$ in line emission at >$3\sigma$ ($8$ are confirmed \cii~emissions at $z$$\sim$$5$ and two are confirmed CO transitions at lower redshifts, see {Loiacono et~al.} \cite{ALPINE_LOIACONO21}).

\subsection{Overview of \textit{ALPINE}'s Research Topics}

\textls[-5]{\textit{ALPINE} is the largest multiwavelength survey to study galaxies in the post-reionization} era at $z=$~4--6. Specifically, the ALMA data are crucial to mitigate common selection biases towards low-dust-attenuated, UV-bright galaxies. It also extends earlier sub-mm samples focusing on intensely star-forming (SFR~$>1000\,{\rm M_{\odot}\,yr^{-1}}$) to main-sequence galaxies. In combination with observations at higher frequencies, the \textit{ALPINE} survey opens the doors to study many different science topics, which we summarize in the following.

The \textit{ALPINE} observations focus on \cii, which is an important cooling line and, therefore, expected to be prevalent in star-forming galaxies. This is especially the case at high redshifts where average SFRs are increased compared to the local universe. Correspondingly, \cii~is related to the SFR, as demonstrated in large samples of low-redshift galaxies (e.g., \citep{CARILLI13,DELOOZE14}). Starting in the early years of ALMA, several studies have shown that it is possible to measure \cii~out to high redshifts, although such studies mostly focused on bright submillimeter galaxies and only small samples of main-sequence \mbox{galaxies~\citep{WILLOTT15,CAPAK15,RIECHERS14,MAIOLINO08}}. The origin of \cii~emission is complicated, which could affect its relation to star formation especially at high redshifts, where star formation is taking place in environments with different chemical and radiation properties compared to low-$z$. While \cii~is often emitted from photo-dissociation regions (PDRs) of molecular clouds, it can also originate from the cold neutral medium and \hii~regions \citep{HERRERACAMUS15,HERRERACAMUS17,PINEDA13,VALLINI15} or traces the diffuse ionized gas \citep{PAVESI16,PAVESI19}. Furthermore, some earlier studies suggested a \cii-deficit (with respect to far-IR luminosity) in high-redshift galaxies (e.g.,  \citep{MAIOLINO15,PENTERICCI16,CARNIANI18}), which could be due to absolutely lower Carbon levels, observations missing broad spatial components of \cii~emission, or simply selection effects. The multiwavelength information (and especially the far-IR continuum emission) from \textit{ALPINE} provides the first verification of the relation between SFR and \cii~emission at these redshifts.

By targeting a large sample of normal main-sequence galaxies, \textit{ALPINE} studies, for the first time, the statistical contribution of dust-obscured star formation in this galaxy population. Serendipitously detected sources put constraints on the remaining fraction of star formation that is not traced by UV-detected galaxies. Further constraints on the dust mass can be placed by the measurement of far-IR continuum emission, and the distribution of dust can be studied by spatial distributions of UV, optical, far-IR continuum light, and \cii~emission or the relation between the $\LIR$/$\LUV$ continuum ratio (also denoted as \textit{IRX}) and the UV continuum slope ($\beta$). The latter \IRXB~diagnostic provides insight into the spatial distribution of dust, dust grain properties, and metallicity \citep{REDDY12,FAISST17b}. In addition, the relation between IRX and stellar mass provides insights into the buildup of dust at this critical epoch of galaxy growth (e.g., \citep{GRAZIANI20}).

At least out to $z$$\sim$2--3, increased gas fractions ({i.e.}, larger gas reservoirs) have been identified as the main reason for the higher SFRs of the galaxies \citep{SCOVILLE17,Tacconi2020}. Commonly, CO lines are good tracers of the molecular gas reservoirs in galaxies. At higher redshifts, only high-$J$ CO transitions can be observed, which come with considerable uncertainties due to necessary excitation corrections.
At higher redshifts, the gas content of main-sequence galaxies has therefore not been studied statistically and robustly. \cii~has been shown to be a reasonable empirical predictor of the molecular gas content, which does not involve uncertainties of excitation corrections associated with high-$J$ CO transitions \citep{HUGHES17,ZANELLA18}. The relation between \cii~and gas mass is likely due to the secondary causal connection with SFR.
With \textit{ALPINE} data, the first statistical assessment of gas masses and fractions can be undertaken via their estimation from \cii. Together with a quantification of the contribution of mergers (derived from close-pair counts in sub-mm data) a full picture of mass growth in $z=$~4--6 galaxies can be obtained.

\cii~is an excellent dust-independent tracer of gas dynamics, which allows studies similar to those undertaken at $z$$\sim$$2$ with \halpha~\citep{FORSTERSCHREIBER09,EPINAT12,MOLINA17}. Specifically, sub-mm observations are sensitive to dust-obscured galaxy mergers and starbursts. \textit{ALPINE} provides the first morpho-kinematic classification of mergers, rotators, and dispersion-dominated galaxies to study the buildup of structure in galaxies as well as the emergence of ``disk galaxies'' ({i.e.}, smooth rotators). Furthermore, the high spectral resolution of ALMA allows to put constraints on outflows and galactic winds via the width and broad components of the \cii~emission line. Such studies have an important impact on our understanding of the baryon cycle ({i.e.}, the outflow, inflow, and recycling of gas) of galaxies at early cosmic times.

Finally, \textit{ALPINE} provides the largest sample of \cii-detections, making it possible for the first time to study the demographics of \cii-line emitters at $z$$\sim$4--6 \citep{ALPINE_YAN20}. More importantly, the calculated \cii-line luminosity functions (LF) allow us to compare with the volume density of \cii-emitters at $z$$\sim$$0$ \citep{HEMMATI17}. The \cii-LF results provide the quantitative constraints on the redshift evolution of \cii-emitter as well as the estimates of molecular gas density at $z$$\sim$4--6. We note the unique aspect of \textit{ALPINE} data, which is that the serendipitous \cii-detections provide the data to compute the incompleteness corrections to the \cii-LF~\citep{ALPINE_LOIACONO21}. 

In Section~\ref{sec:results}, we summarize the results of \textit{ALPINE} on the topics above in more detail.

\section{Results}\label{sec:results}

\subsection{\cii~and Far-Infrared Continuum Luminosity Function} \label{sec:luminosityfunctions}

The \cii~line and far-IR continuum emission are the primary ALMA measurements for the \textit{ALPINE} galaxies. These measurements are obtained for the main \textit{ALPINE} targets as well as for the galaxies that are serendipitously detected in far-IR continuum and \cii~line emission in a total area of $24.92\,{\rm arcmin^2}$ (referred to as the \textit{ALPINE} ``blind survey''). In the following, we summarize the results on the total far-IR and \cii~luminosity functions obtained from these basic measurements. For more detailed descriptions, we refer the reader to the original \textit{ALPINE} papers on these topics \citep{ALPINE_BETHERMIN19,ALPINE_GRUPPIONI20,ALPINE_LOIACONO21,ALPINE_YAN20}.

The average continuum and \cii~line $1\sigma$ sensitivity reached by \textit{ALPINE} are $29\,\upmu{\rm Jy}$ and $0.14\,{\rm Jy\,km\,s^{-1}}$, respectively. At $3.5\sigma$, this yields 23 detections in the continuum and 75~detections in \cii~emission, which correspond to $19.5\%$ and $63.5\%$, respectively, of the total $118$ targeted galaxies (see \citep{ALPINE_BETHERMIN19}). Within the same fields of the primary targets, $56$~continuum and $12$ \cii~line emitters are serendipitously detected. These detections are reported in \citet{ALPINE_GRUPPIONI20} and \citet{ALPINE_LOIACONO21}.

The 56 serendipitous far-IR continuum sources are identified using the \textit{AstroPy} routine \texttt{find\_peaks} in the signal-to-noise (S/N) ratio maps by applying a threshold of ${\rm S/N}=5$ (resulting in a purity of $95\%$, \citep{ALPINE_BETHERMIN19}). S/N maps are used to mitigate the issues of variable sensitivity across the ALMA FoV. The redshifts of the $56$ sources are subsequently constrained by matching their positions to ancillary catalogs as well as fitting their SEDs with \textsc{LePhare}\endnote{\url{http://www.cfht.hawaii.edu/~arnouts/lephare.html}, accessed on May 31, 2022} \citep{ARNOUTS99,ILBERT06}. Of the 56 serendipitous continuum sources, at least 47 have photometric redshifts, 5 have spectroscopic redshifts from \cii~emission lines, and 4 have no redshifts. We note that $6$ out of the $56$ sources are only detected in ALMA (hence, they are ``optically dark'') and their photometric redshifts are therefore highly uncertain.
The final redshift distribution of the $52$ sources ranges from $z$$\sim$$0.5$ to $z$$\sim$$6$ with a peak at a median redshift of $\langle z \rangle = 2.84 \pm 0.18$ \citep{ALPINE_GRUPPIONI20}. Interestingly, 10 sources are found at $z$$\sim$4--6, potentially due to physical overdensities around the main \textit{ALPINE} targets, suggesting abundant clustering of galaxies at $z$$\sim$$5$.

The serendipitous \cii~emitters are identified on the full three-dimensional data cube using the \texttt{find\_clumps} algorithm \citep{DECARLI16,WALTER16} by applying a threshold of ${\rm S/N}>6.3$, which corresponds to a fidelity (proportional to the ratio of negative to positive peaks) of >$85\%$. Two additional line emitters (at ${\rm S/N}$$\sim$$5.98$) with optical counterparts are added due to their S/N being close to the threshold. In total, $14$ serendipitous line emitters are subsequently matched to photometric and spectroscopic catalogs.
Out of these $14$ candidates, $8$ have redshift confirming \cii~emission at $4.3 < z < 5.4$. Specifically, $4$ of them have spectroscopic redshifts and $4$ have photometric redshifts (consistent with \cii~emission) based on the \textit{COSMOS} multiwavelength photometry, including UV to near-IR bands (with photo-z uncertainties $\Delta z<0.2$; \citep{LAIGLE16}). Of the remaining candidates, two galaxies are identified as high-$J$ CO transition at $z=1.3$ and $z=0.9$, and the identification of $4$ lines is ambiguous based on the available photometry.

\subsubsection{The \cii~Luminosity Function}\label{sec:cii_lf}

The \cii~luminosity function is derived by a $V_{\rm max}$ method \citep{SCHMIDT68} using the \textit{ALPINE} main target galaxies as well as the serendipitously detected \cii~emitters. The inclusion of the latter is important to mitigate selection biases in the main \textit{ALPINE} target sample. Both luminosity functions \citep{ALPINE_YAN20, ALPINE_LOIACONO21} are first corrected for incompleteness. The serendipitous sample additionally includes a correction factor based on the fidelity of the serendipitous detections. The completeness correction of the target sample is based on the fact that the \textit{ALPINE} sample builds a subset of UV-detected galaxies \citep{ALPINE_YAN20}. For the serendipitous sample, the completeness has been computed using injection simulations \citep{ALPINE_BETHERMIN19,ALPINE_LOIACONO21}.

\begin{figure}[t]
\includegraphics[width=0.5\textwidth]{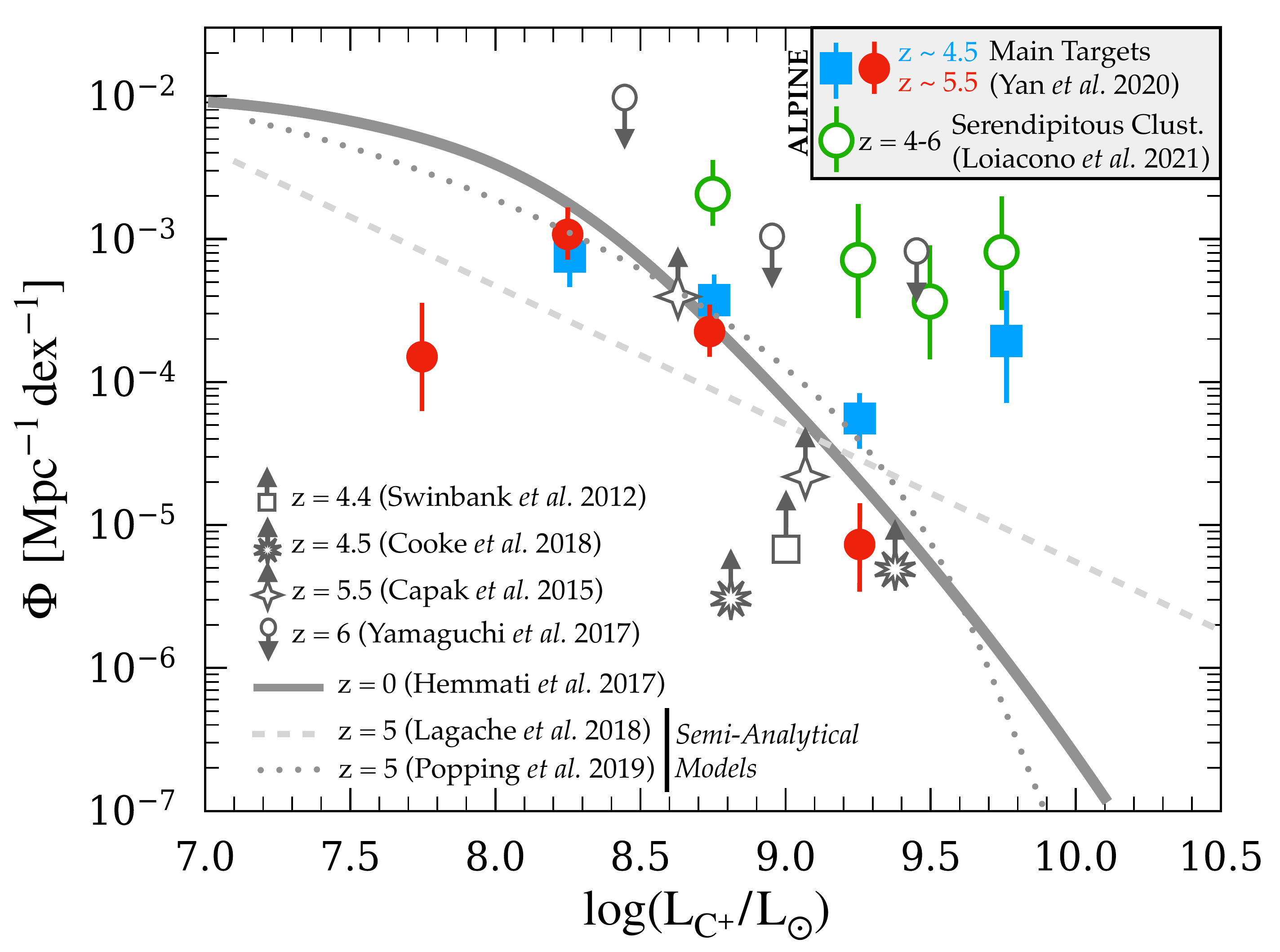}
\includegraphics[width=0.5\textwidth]{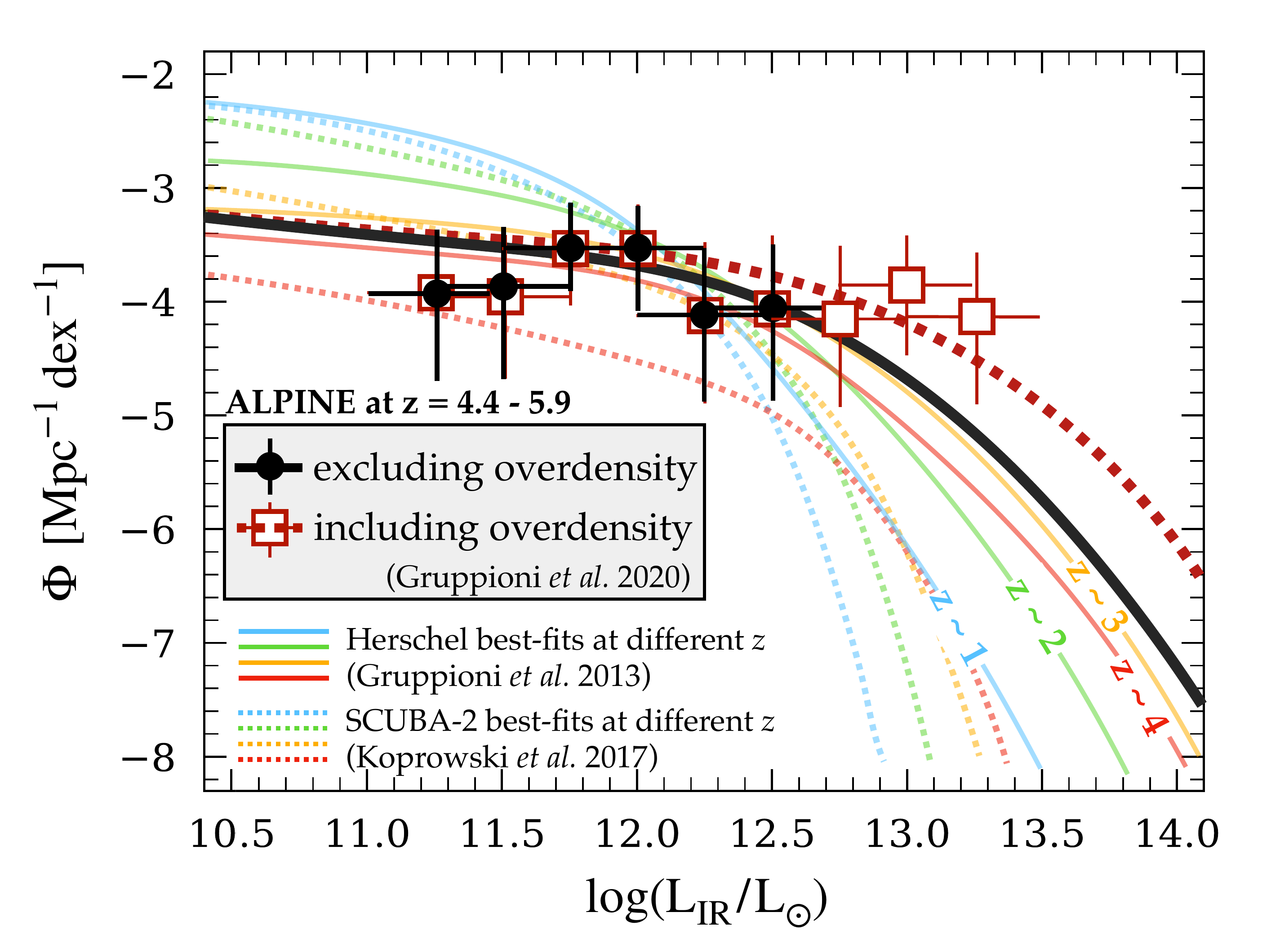}
\caption{\cii~line emission and total far-IR LFs from \textit{ALPINE} and literature. \textit{Left Panel:} \cii~LF measured for the \textit{ALPINE} main sample (blue squares, $z$$\sim$$4.5$; red circles, $z$$\sim$$5.5$; Yan et~al. 
\cite{ALPINE_YAN20}) as well as serendipitous sample (incl. clustered galaxies) from the \textit{ALPINE} blind survey (green open circles; \citep{ALPINE_LOIACONO21}).
Note that the counts of serendipitous \cii~emitters may be elevated due to clustering (see details in \citep{ALPINE_LOIACONO21}). Other results from the literature \citep{SWINBANK12,CAPAK15,YAMAGUCHI17,COOKE18} at $z$ = 4.4, 4.5, 5.5, and $6$ are also shown with respective symbols (mostly limits). For comparison, the $z$$\sim$$0$ \cii~LF from~\mbox{\citet{HEMMATI17}} is shown (solid gray line) as well as predictions from semianalytical models (gray dashed and dotted lines; \citep{LAGACHE18,POPPING19}). Note that the \textit{ALPINE} data become significantly incomplete and uncertain at $\LCII < 10^{8.5}\,L_{\odot}$ due to survey limits and the uncertain incompleteness correction because of the lack of blindly detected \cii~samples (serving as incompleteness correction sample for the main targets). Hence, LF is the lower limit.
 Figure adapted from \citet{ALPINE_YAN20} and \citet{ALPINE_LOIACONO21}.
\textit{Right Panel:} Total far-IR LF from \textit{ALPINE} continuum detections at $z$ = 4.5--5.9 \citep{ALPINE_GRUPPIONI20} excluding overdensities (black circles) and including overdensities (open red squares). The solid, thick black line and dotted red lines, respectively, show the best-fit Schechter functions. The best fits to measurements from Herschel (thin solid lines; \citep{GRUPPIONI13}) and SCUBA-2 S2CLS (thin dotted lines; \citep{KOPROWSKI17}) for different redshifts (from $z=1$ to $z=4$; from blue to red) are also shown. Adapted from Ref. \cite{ALPINE_GRUPPIONI20}.
\label{fig:lf}}
\end{figure}

The left panel in Figure~\ref{fig:lf} shows the resulting luminosity function at $z$$\sim$$5$ in comparison with the \cii~LF at $z$$\sim$$0$ (solid gray line from \citet{HEMMATI17}) as well as other literature measurements at higher redshifts.
It is important to note that the \cii~LF at $z$$\sim$$5$ derived from the \textit{ALPINE} main sample is a lower limit, as it does not include serendipitously detected \cii~emitters, which can add to the number counts (see \citep{ALPINE_LOIACONO21}). The incompleteness correction of the LF is very uncertain as it depends on the serendipitous sample, which is too small and covers only a narrow luminosity interval (mostly high-luminosity end at $\LCII>10^{8.5}\,L_{\odot}$). In addition, the serendipitous \cii~emitters seem to be clustered with the main targets (see below); therefore, they are not representative of the field population, which adds additional uncertainty. Furthermore, the \textit{ALPINE} data are not able to constrain the \cii~LF at $\LCII < 10^{8}\,L_{\odot}$ due to the survey limits.
Given these caveats, the $z$$\sim$$5$ \cii~LF ({i.e.},  volume density of \cii-emitters per magnitude interval) seems to be comparable with that of $z$$\sim$$0$ at $\LCII = 10^{8.5}\text{--}10^{9.5}\,L_{\odot}$.
This result is in stark contrast with the strong evolution observed in the rest-frame UV LFs \citep{Ono2018,Bouwens2007}.
At the high-luminosity end (>$10^{9.5}\,L_\odot$), a 1--2~$\sigma$ significant excess of \cii~emitters compared to the local \cii~luminosity function \citep{HEMMATI17} is found (Figure~\ref{fig:lf}). The excess compared to $z=0$ is amplified by the inclusion of ``UV dark'' \cii~emitters (serendipitously detected and missed by the \textit{ALPINE} main sample). Clustering of \cii~emitters also plays a role in enhancing the number density at bright luminosities. This is shown by the green symbols in the left panel of Figure~\ref{fig:lf}, which include serendipitous sources close to the main targets.

The \cii-LF result is also compared with that of CO LFs at $z$$\sim$3--5.8 derived by ASPECS~\citep{DECARLI19} and COLDz \citep{RIECHERS19} using the CO($1-0$) to \cii~conversion relation (see Equations~(5) and (6) in \citet{ALPINE_YAN20}). The CO LFs confirm the excess of luminous \cii-emitters that were found in the \textit{ALPINE} data, suggesting that at $\rm \LCII > 10^9\,L_\odot$ UV-faint but \cii-bright sources likely make significant contributions to the \cii~emitter volume density. 
The derived \cii-LFs, in combination with far-IR and CO LFs, are also compared with current available model predictions \citep{POPPING16,LAGACHE18}, which are found to significantly underestimate the number densities of \cii~emitters at $z$$\sim$4--6. 

\subsubsection{The Total Far-Infrared Luminosity Function}\label{sec:fir_lf}

The computation of the total far-IR luminosity is less straightforward (see details in~\citep{ALPINE_GRUPPIONI20}). As ALMA only measures one continuum point at rest-frame $\sim$$150\,\upmu{\rm m}$ for the \textit{ALPINE} sample, a far-IR SED shape has to be assumed in order to obtain the total far-IR luminosity (integration between rest-frame 8--1000 $\upmu{\rm m}$).
To this end, an average far-IR SED is created by stacking \textit{Herschel} measurements of several hundred $z=$~4--6 galaxies on the \textit{COSMOS} field in the same stellar mass and SFR range as the \textit{ALPINE} sample (${\rm SFR_{UV}} > 10\,{\rm M_{\odot}\,yr^{-1}}$)  \citep{ALPINE_BETHERMIN19}. The obtained far-IR SED template is similar to a modified black body with a luminosity-weighted dust temperature of $42\,{\rm K}$, and is also in agreement with fits to $4$ \textit{ALPINE} galaxies with multiple far-IR continuum measurements (see \citet{FAISST20b}). This SED template will also be used later to derive infrared-based SFRs of the \textit{ALPINE} galaxies from the total far-IR luminosity.

For the derivation of the total far-IR LF across cosmic time, the $56$ serendipitous far-IR continuum sources from the \textit{ALPINE} blind survey are used in order to mitigate potential selection biases towards strong \cii~emitters. Similar to the derivation of the \cii~LF, a $V_{\rm max}$ method is used to obtain the far-IR luminosity function. The resulting luminosity functions (right panel of Figure~\ref{fig:lf}) from $z$ = 0.5--6 (differentiated in colors) are similar in shape with no significant decrease in their normalizations. The results are also in good agreement with those derived from the CO luminosity functions \citep{RIECHERS19,DECARLI16,DECARLI19}.

All in all, there is still a large uncertainty in the derived total far-IR LFs due to small sample sizes and selection effects (for example, changes in galaxy densities at small and large scales add significant uncertainties to the number counts as shown by \citet{ALPINE_LOIACONO21} and \citet{ALPINE_GRUPPIONI20}). Large blind surveys would be a way to remedy these uncertainties;, however, unless very deep, they are usually only sensitive to the most-luminous sources and do not cover the full luminosity range.

\subsection{Is \cii~a Good Tracer of Star Formation at $z>4$?}\label{sec:ciisfr}

A correlation between \cii~emission and total SFR is expected from studies crossing cosmic time from $z=0$ to $z$$\sim$$2$ \citep{PINEDA14,HERRERACAMUS15,DELOOZE11,DELOOZE14}. Such a relation is fundamental as \cii~is one of the strongest far-IR lines and is now widely observed in high-$z$ galaxies. The far-IR continuum (a more-common SFR indicator; \citep{KENNICUTT98}) is significantly fainter; hence, it is much more time-consuming to detect. While probed by large datasets at local redshifts, the \cii$-$SFR relation at higher redshifts is uncertain. Specifically, several studies suggest a population of \cii underluminous galaxies due to their observed low \cii/$\LFIR$ ratios compared to lower \mbox{redshifts \citep{MAIOLINO15,WILLOTT15,PENTERICCI16,BRADAC17,CARNIANI18,CAPAK15,RIECHERS14}.} The origin of \cii~from various regions in the galaxy (\hii~regions, diffuse neutral and ionized ISM and PDRs; \citep{WOLFIRE95,HOLLENBACH99}) add additional~complications. 



One of the main goals of \textit{ALPINE} is to revisit the \cii$-$SFR relation at $z$$\sim$$5$ using the largest sample of \cii~emitters to date and compare it to observations at lower redshifts as well as theoretical models. The full analysis is detailed in \citet{ALPINE_SCHAERER20}. Here, the most important results are summarized.

As discussed in Section~\ref{sec:luminosityfunctions} and further elaborated in Section~\ref{sec:sfrd_dust}, a significant population of dust-attenuated and -obscured galaxies has been suggested even at \mbox{$z>4$}. As pointed out in \citet{ALPINE_SCHAERER20}, an accurate assessment of the total SFR ({i.e.}, ${\rm SFR_{UV}}+{\rm SFR_{FIR}}$) and a robust inclusion of limits in \cii~and far-IR continuum is therefore crucial for such an analysis.
The total SFR of the \textit{ALPINE} main targets are derived in \mbox{\citet{ALPINE_SCHAERER20}} as follows.
First, the dust uncorrected UV SFRs (${\rm SFR_{UV}}$) are based on the rest-frame $L_{\rm 1500}$ luminosity derived from SED fitting at $1500\,{\rm \A}$ (see details in \citet{ALPINE_FAISST20}) and by applying the \citet{KENNICUTT98} relation, ${\rm SFR_{UV}} [{\rm M_{\odot}\,yr^{-1}}]~=~1.59 \times 10^{-10}\,L_{\rm 1500}/{\rm L_{\odot}}$ converted to a \citet{CHABRIER03} IMF.
For the calculation of the infrared SFR (${\rm SFR_{FIR}}$), the conversion from ALMA-measured single-point photometry at rest-frame $\sim$$150\,\upmu{\rm m}$ to total far-IR luminosity ($L_{\rm IR}$) is crucial. The conversion depends strongly on the assumed shape of the far-IR SED (itself depending on many parameters including the warm and cold dust temperature), which is unknown as only one far-IR data point exists for the \textit{ALPINE} galaxies. To compute the conversion factor, an average empirical far-IR template is created using the same approach as outlined in Section~\ref{sec:fir_lf} (for details, see \citep{ALPINE_BETHERMIN19}).
The far-IR SFRs for far-IR-continuum-detected galaxies are derived via ${\rm SFR_{FIR}} [{\rm M_{\odot}\,yr^{-1}}] =  1.40 \times 10^{-10}\,L_{\rm IR}/{\rm L_{\odot}}$ \citep{KENNICUTT98}, where the far-IR luminosity is again obtained by integration of the far-IR SED at 8--1000 ${\upmu \rm{m}}$. For galaxies not detected in far-IR continuum ({i.e.}, ALMA sub-mm), $L_{\rm IR}$ is derived from the stacked \IRXB~relation\endnote{The \IRXB~relation relates the rest-UV continuum slope ($\beta$, described in \citet{ALPINE_FAISST20}) to the $L_{\rm IR}/L_{\rm UV}$ (IRX) luminosity ratio.} presented in \citet{ALPINE_FUDAMOTO20} (see also Section~\ref{sec:sfrd_dust}).

\begin{figure}[t]
\includegraphics[width=0.85\textwidth]{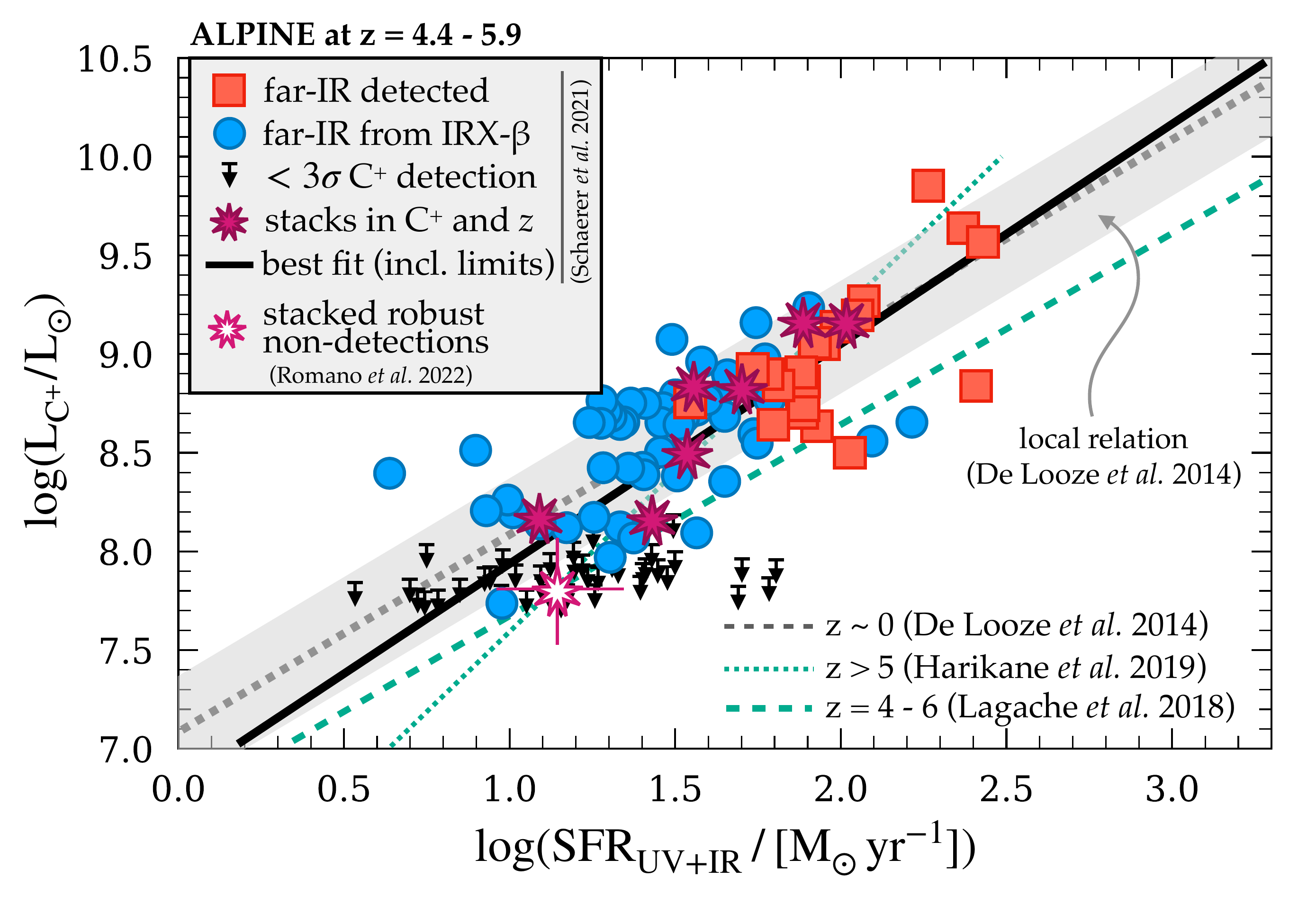}
\caption{\cii~vs. total
 SFR (from UV + far-IR) relation derived from the \textit{ALPINE} sample from
 \citet{ALPINE_SCHAERER20}. Galaxies detected in far-IR are shown as red squares. For galaxies not detected in far-IR (blue circles), the far-IR luminosity is derived from the \IRXB~relation from \citet{ALPINE_FUDAMOTO20}. \cii~non-detections at $3\sigma$ are shown as black upper limits. The filled purple stars show stacked measurements (four bins of $\LCII$~and two bins in redshift). The empty purple star shows a stack from robust nondetections from \citet{ALPINE_ROMANO22}. The best fit (${ \rm log(L_{C^+}/L_{\odot})} = \alpha + \beta \times {\rm log(SFR / [M_{\odot}\,yr^{-1}])}$ with $\alpha = 6.61\pm0.20$ and $\beta=1.17\pm0.12$) is shown as a thick black line. The $z=$~4--6 relation from \textit{ALPINE} is consistent with the relation derived from local galaxies (gray dashed line with shaded region ($2\sigma$);~\citep{DELOOZE14}).
The predictions from the semianalytical model from \mbox{\citet{LAGACHE18}} (green dashed line) and the observations at $z>5$ from \citet{HARIKANE18} (green dotted line) are also shown.
Adapted from Ref. \cite{ALPINE_SCHAERER20}.
\label{fig:ciiSFR}}
\end{figure}

Figure~\ref{fig:ciiSFR} shows the resulting relation between \cii~and SFR as measured from the \textit{ALPINE} sample at $z=$~4--6 from \citet{ALPINE_SCHAERER20}.
The best-fit relation is shown as a solid black line and is derived using the Bayesian fitting method described in \citet{KELLY07}\endnote{The method is implemented in the Python package \textit{linmix} at \url{https://github.com/jmeyers314/linmix}, accessed on May 31, 2022.}, which properly treats errors and nondetections in astronomical data during the fitting process.
We also show the location of the stacked robust nondetections from (purple empty star \citet{ALPINE_ROMANO22}), which agrees well with the \cii~vs. SFR relation found by the stacked detections (purple filled stars). 
In addition, the observed \cii~vs. SFR relation of local galaxies (gray line and region ($2\sigma$); \citep{DELOOZE14}) as well as theoretical models \citep{LAGACHE18}\endnote{Note that all of these relations have been calibrated to the same IMF.} are also shown. While other studies found \cii~to be underluminous \citep{MAIOLINO15,WILLOTT15,PENTERICCI16,BRADAC17,CARNIANI18}, the \textit{ALPINE} sample shows a very consistent relation between \cii~and total star formation, similar to findings at lower redshifts \citep{DELOOZE14}. Even more so, the comparison of \textit{ALPINE} data with measurement from the literature at higher redshifts \citep{MATTHEE19,HARIKANE18} suggests only a slow (if any) evolution of the \cii$-$SFR relation out to higher redshifts.

\subsection{SFR Density, Main-Sequence, and Dust Abundance across Cosmic Time}\label{sec:sfrd_dust}

The data presented Section~\ref{sec:luminosityfunctions} suggest that a significant fraction of the UV light is obscured in main-sequence galaxies. In this section, we present more detailed results on the dust abundance and the implication on the total cosmic SFR density.

\begin{figure}[t!]
\includegraphics[width=0.5\textwidth]{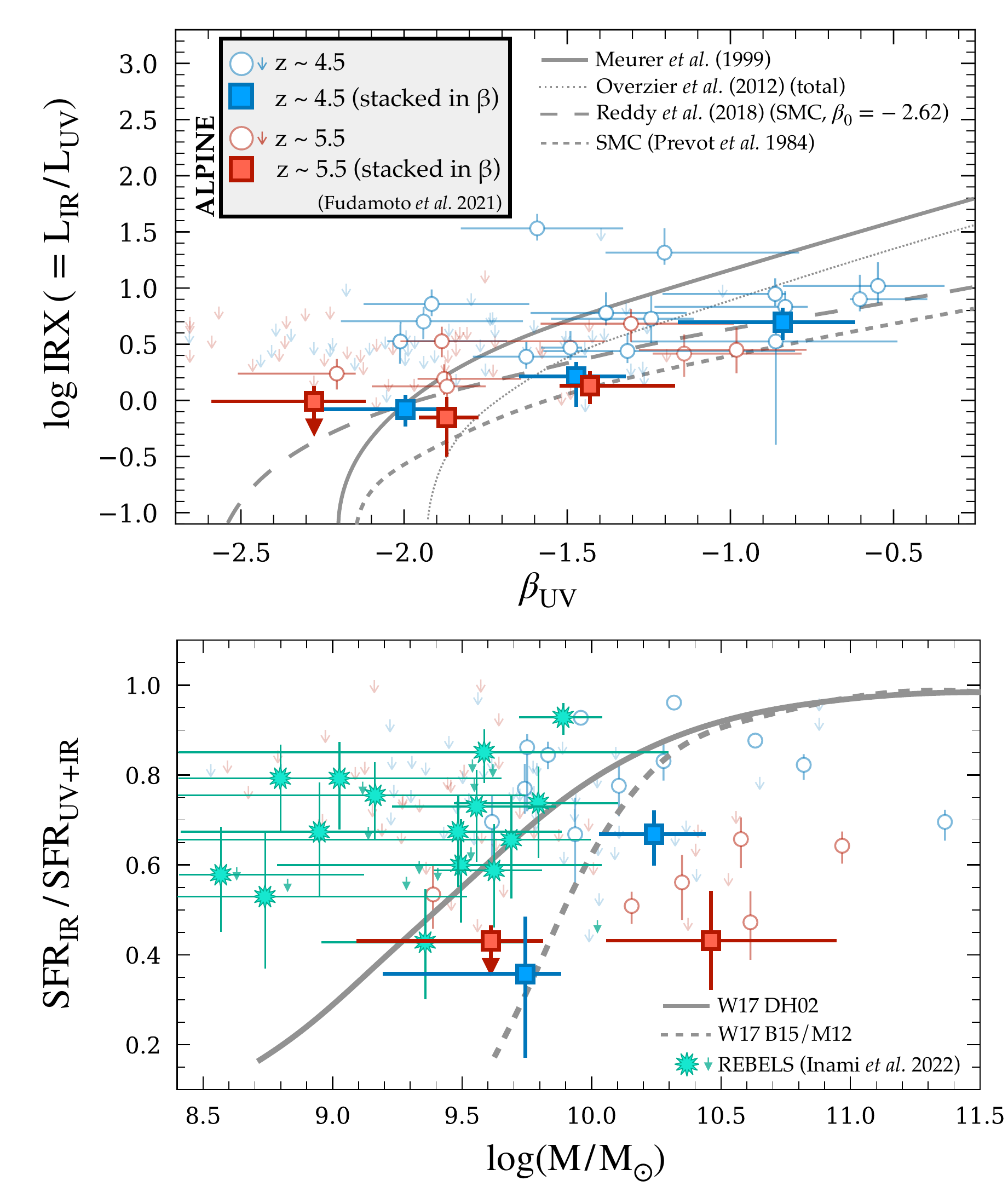}
\includegraphics[width=0.5\textwidth]{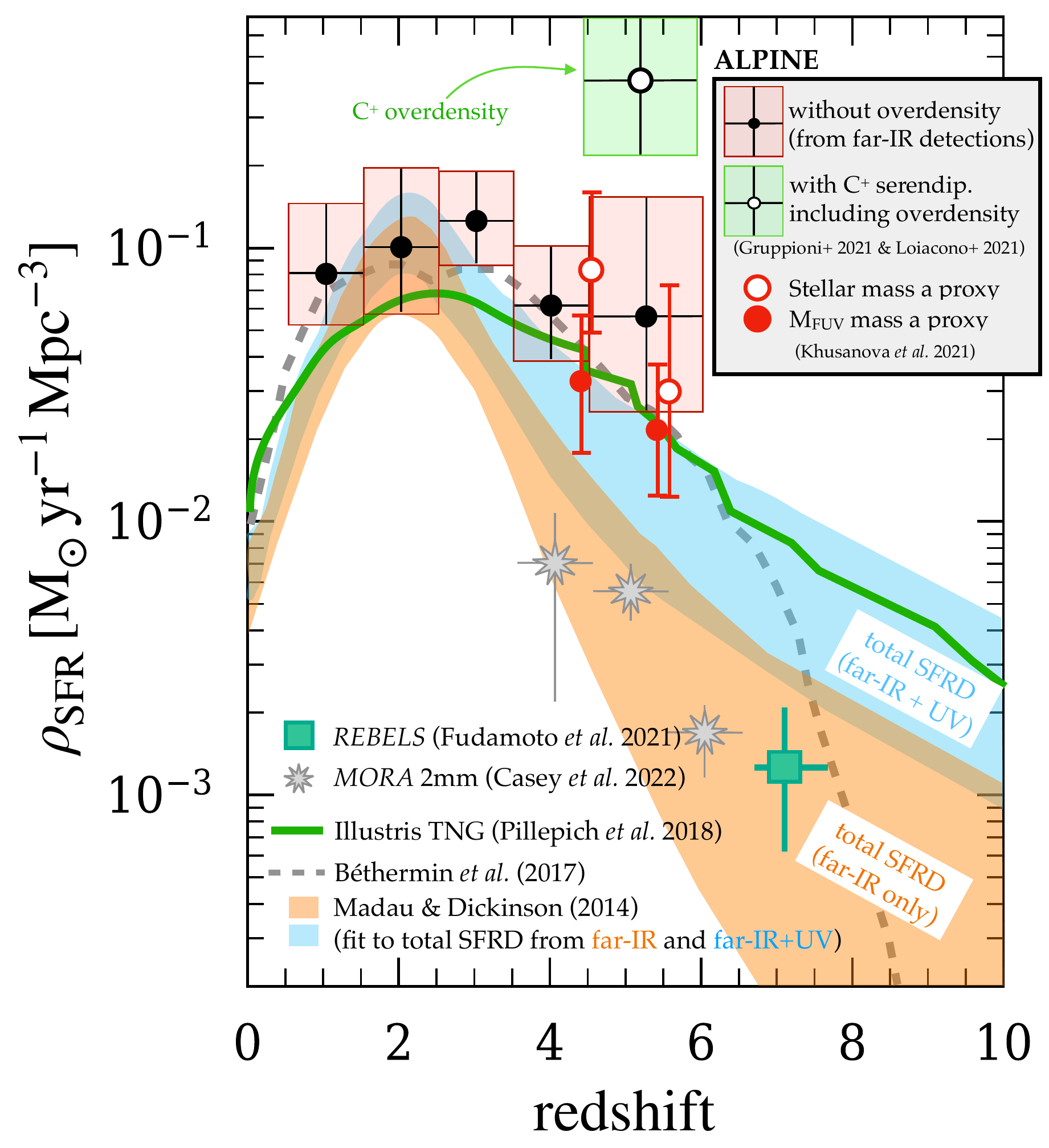}
\caption{Dust obscuration and total SFR density from the \textit{ALPINE} sample.
\textbf{Left Panels:} \IRXB~relation (\textit{top}) and total obscured fraction of SFR as a function of stellar mass (\textit{bottom})
 \citep{ALPINE_FUDAMOTO20}. Individual \textit{ALPINE} galaxies at $z$$\sim$$4.5$ and $z$$\sim$$5.5$ are indicated in blue and red circles (and limits). Stacked measurements in bins of $\beta$ and stellar mass, respectively, are shown with filled squares. On the \IRXB~diagram, we also show SMC-like dust (short-dashed line; \citep{PREVOT84}), the starburst relation (solid line; \citep{MEURER99}) and the updated local relation (dotted line; \citep{OVERZIER11}), and the relation by \citet{REDDY18} for $\beta_0=-2.62$ (long-dashed line). On the obscured fraction vs. stellar mass diagram, we also show the \textit{REBELS} sample (only with ALMA redshift, teal stars; \citep{INAMI22}) and the $250\,{\upmu \rm{m}}$ data from \citet{WHITAKER17} at $z$ = 0--2.5 using the far-IR template from \citet{DALE02} (solid line) and \mbox{\citet{BETHERMIN15}} or \mbox{\citet{MAGDIS12}} (dashed line) for the calculation of the total SFR. Figures adapted from \mbox{\citet{ALPINE_FUDAMOTO20}}.
\textbf{Right Panel:} Total SFR density as a function of redshift derived from the \textit{ALPINE} blind sample (black circles and red squares; \citep{ALPINE_GRUPPIONI20}) and stacked far-IR measurements (red symbols; \citep{ALPINE_KHUSANOVA20}). We also include the result from serendipitous \cii~detections that show a higher density because of clustering with the \textit{ALPINE} main targets \citep{ALPINE_LOIACONO21}. Measurements from the \textit{MORA} 2mm ALMA survey (far-IR SFRs only) from \citet{CASEY21} are shown as gray stars. The predictions from Illustris TNG (green line \citep{PILLEPICH18}) and \citet{BETHERMIN17} (dashed line) are also shown. The total SFR density fits for ${\rm SFR_{IR}}$ and ${\rm SFR_{UV+IR}}$ data from the \citet{MADAU14} review are shown as orange and blue swaths, respectively. Adapted from Ref. \cite{ALPINE_GRUPPIONI20}.
\label{fig:irx_sfrd}}
\end{figure}  

The fraction of dust-attenuated and -obscured star formation in the \textit{ALPINE} sample is studied by \citet{ALPINE_FUDAMOTO20} using two diagnostic diagrams, namely, the \IRXB~and the \IRXM~relations.
The measurements of the UV continuum slope ($\beta$), UV luminosity, and stellar mass are described in detail in \citet{ALPINE_FAISST20}. In brief, the quantities are determined from the UV to optical best-fit SED to the broad and narrowband photometry of the galaxies. The $\beta$ slope ($f_{\lambda} \propto \lambda^{\beta}$) is measured between a rest-frame wavelength of $1300\,{\rm \A}$ and $2300\,{\rm \A}$, consistent with previous studies (e.g., \citep{BOUWENS16,MCLURE18}). The total far-IR luminosity was determined from the ALMA continuum data by using the SED template derived in \citet{ALPINE_BETHERMIN19} (see also Section~\ref{sec:ciisfr}).
The ALMA far-IR continuum images of the \textit{ALPINE} galaxies are then stacked in bins of $\beta$ and stellar mass to obtain an average of the \IRXB~and \IRXM~relations.

The left panels of Figure~\ref{fig:irx_sfrd} show the \IRXB~relation as well as a slightly modified version of the \IRXM~relation, namely, the relation between the fraction of dust-obscured SFR and stellar mass, for two different redshifts ($z$$\sim$$4.5$ and $z$$\sim$$5.5$).
This analysis shows that the \IRXB~relation obtained from \textit{ALPINE} is generally below the \citet{MEURER99} relation, which is based on local starburst galaxies, suggesting a steepening of the attenuation curve. This confirms suspicions of earlier studies based on samples that are a factor of $10$ and~smaller~\citep{CAPAK15,BARISIC17}. Furthermore, the obscured fraction of star formation is significantly lower ($65\%$ and $45\%$ at $z$$\sim$$4.5$ and $z$$\sim$$5.5$, respectively) compared to galaxies at $z<3$ ($80\%$) at high masses of 1--3 $\times 10^{10}\,{\rm M_{\odot}}$. For lower stellar masses, the obscured fractions are statistically consistent. 
Altogether, this suggests a rapid evolution of dust attenuation and dust properties in massive (>$10^{10}\,{\rm M_{\odot}}$) galaxies between $z$$\sim$$6$ and the peak of cosmic SFR density at $z$$\sim$2--3.
Specifically, a steeper dust attenuation curve could be a sign of dust produced by supernovae (SNe) \citep{MAIOLINO04,HIRASHITA05,STRATTA07,GALLERANI10}.
Such a scenario is favored in young, star-forming galaxies due to the lack of time for other dust production mechanisms to work (e.g., dust production from Asymptotic Giant Branch (AGB) stars or dust growth in the ISM \citep{ASANO13,DWEK11,TODINI01,NOZAWA03,SCHNEIDER04}). 
However, further investigations are crucial to definitively decipher the dust production mechanisms at these cosmic times.
Further, it is interesting that there is a large range of dust attenuation values at $z=5$. This suggests a wide range of dust production or geometrical effects.

The total cosmic SFR density is an important tool to constrain the evolution of galaxies. It has been measured robustly below $z$ = 2--3, where multiwavelength data (especially in the far-IR) are available. At higher redshifts, the evolution of the SFR density is more uncertain due to the lack of far-IR data to measure the total SFR and to include dust-obscured galaxies.
\citet{ALPINE_KHUSANOVA20} derives the first robust measurement of the total cosmic SFR density at $z=$~4--6 based on UV$+$ALMA multiwavelength data from \textit{ALPINE}. The detailed procedure of this measurement is described in the respective paper. In brief, the total SFR density is obtained by integrating over the full SFR function, which was derived from two different proxy density distribution functions: the absolute UV ($M_{\rm UV}$) LF and the stellar mass function. The total SFR density is then obtained by integration from 0.03--100~${\rm L_{*}}$ or $\logm$ = 6.0--12.4.
The right panel in Figure~\ref{fig:irx_sfrd} shows the resulting SFR density function derived for the \textit{ALPINE} galaxies (red symbols). The red boxes show the derived SFR density from serendipitous far-IR detections from \citep{ALPINE_GRUPPIONI20}. The green box shows the SFR density when including an overdensity of serendipitously detected \cii~emitters~\citep{ALPINE_LOIACONO21}. The \textit{ALPINE} measurements are slightly higher than the prediction from Illustris TNG \citep{PILLEPICH18} and measurements from \citet{MADAU14}, but in line with the measurements from \citet{BETHERMIN17}. The contribution of galaxies to the dust-obscured SFR density from the \textit{MORA} $2\,{\rm mm}$
survey \citep{CASEY21} are also shown.

\subsection{Gas Reservoirs and Baryon Cycle at $z>4$} \label{sec:gas_baryoncycle}

The measurement of the molecular gas content (or gas fraction, $M_{\rm molgas} / (M_{\rm molgas} + M_{*})$) of galaxies as a function of cosmic time is crucial to understand their progress in forming stars. Specifically, the 5--10 times higher specific SFRs (sSFRs) at early cosmic times compared to local galaxies \citep{MADAU14} require either more substantial gas reservoirs or a higher star-formation efficiency in these galaxies. At the same time, vigorous star formation is connected with outflows that enrich the surrounding intergalactic medium (IGM) with processed chemical elements.

\begin{figure}[t!]
\includegraphics[width=1\textwidth]{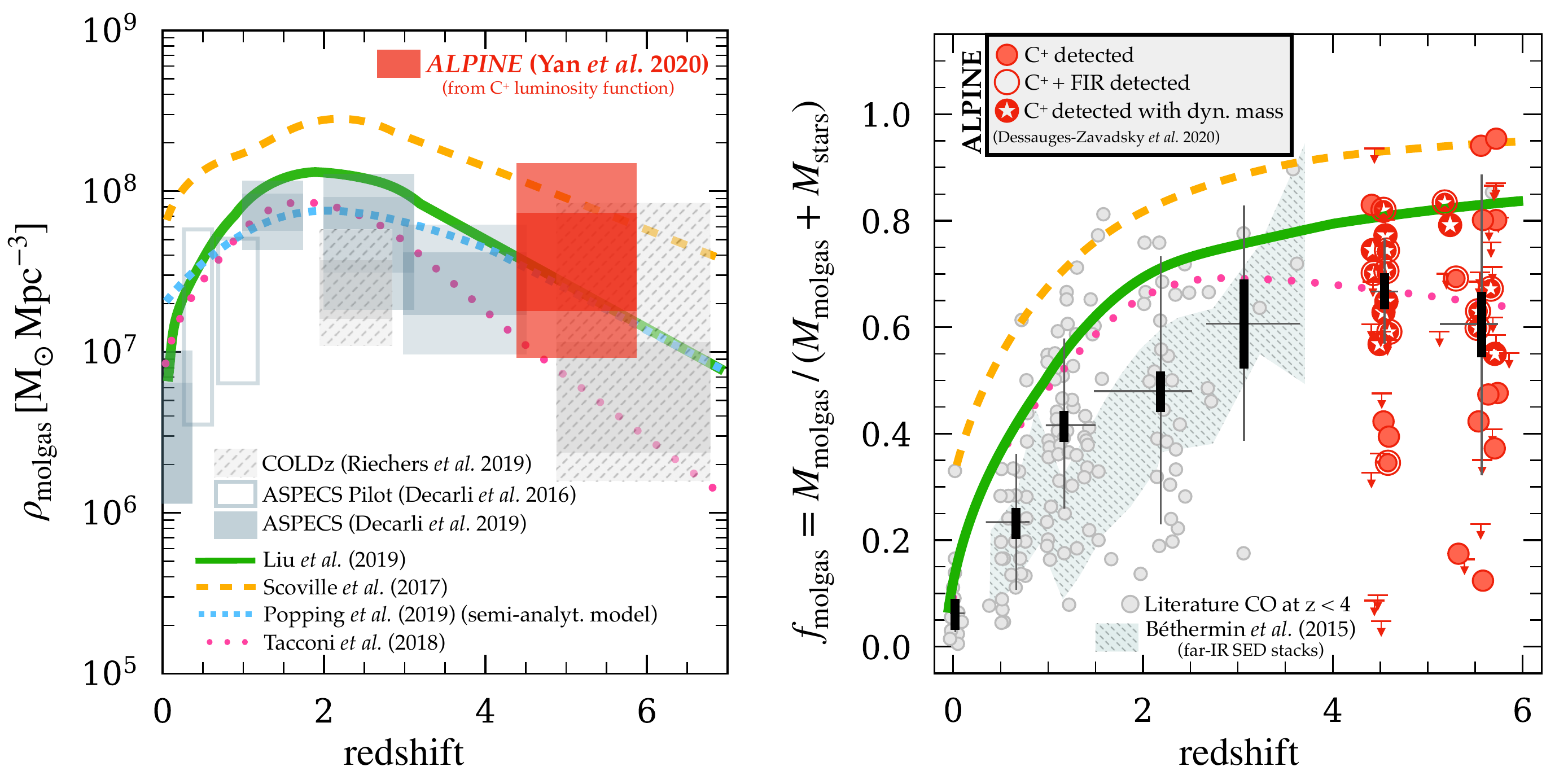}
\caption{\textbf{Left Panel:} Molecular gas density as a function of redshift (following closely the cosmic SFR density, see Figure~\ref{fig:irx_sfrd}). The measurement from \textit{ALPINE} from Yan et~al. 
 \cite{ALPINE_YAN20} is indicated by the red box. Other measurements from \textit{COLDz} (hatched boxes; \citep{RIECHERS19}) and \textit{ASPECS} (filled and empty gray boxes; \citep{DECARLI16,DECARLI19}), together with empirical models from \citep{LIU19} (green solid), \citet{SCOVILLE17} (orange long-dashed), and \citet{TACCONI18} (pink dotted) 
 are also shown. The semi-analytical model from \citet{POPPING19} is shown as a short-dashed blue line.
\textbf{Right Panel:} Redshift evolution of molecular gas mass fraction of galaxies from $z$ =0--6. The curves show the same empirical and semianalytical models as indicated on the left panel. Measurements at $z=4.6$ from \textit{ALPINE} \citep{ALPINE_DESSAUGES20} are shown in red (distinguishing from \cii~detected, far-IR detected, and measurements with dynamical mass). Literature CO data at $z<4$ are shown as gray points (see \citet{ALPINE_DESSAUGES20} for references), together with the far-IR SED stacks from (hatched region \citet{BETHERMIN15}). Large black crosses show the median (with error on the mean and scatter) in $7$ redshift bins.
\label{fig:gas_baryons}}
\end{figure}   

\subsubsection{Gas Fractions, Depletion Times, and Large Range in Gas Properties}

The gas content of galaxies has been measured commonly via CO emission (a tracer of H$_2$) out to $z$$\sim$$3$~\citep{CARILLI13} and it is found to increase with redshift, reflecting the increase in sSFR~\citep{MADAU14}. The measurement of the gas content of galaxies out to higher redshift is challenging and has not been measured for large samples of typical galaxies prior to \textit{ALPINE}. Low-$J$ CO emission becomes increasingly more difficult to observe at high redshifts because of surface brightness dimming, lower metal content of the galaxies, and redshift itself. Furthermore, the increasing cosmic microwave background temperature with redshift decreases the contrast of the line \citep{DACUNHA13}. On the other hand, the conversion of high-$J$ CO transitions to CO(1-0) requires knowledge of the CO excitation stage and the gas density. In addition, even with state-of-the-art millimeter facilities such as \textit{VLA} or \textit{NOEMA}, a detection of high-$J$ CO emission becomes unfeasible for a large sample of galaxies.
For understanding the molecular gas content of high redshift galaxies, other tracers of molecular gas have to be explored. Among these are the dust mass inferred from an SED fit to the far-IR emission~\citep{DALE02,Sajina2006,Casey2012}, the cold dust continuum measured in the Rayleigh--Jeans (RJ) tail regime at $\lambda_{\rm rest} \approx 850\,{\upmu \rm{m}}$ \citep{SCOVILLE14,SCOVILLE17}, and the empirically measured relation between \cii~and molecular gas mass \citep{ZANELLA18,LIU19,MADDEN20}. Furthermore, using an estimate of stellar, dust, and dark matter mass inside a given radius, dynamical mass estimates from kinematics can be used to constrain gas masses \citep{Solomon2005,GENZEL15}.
The \textit{ALPINE} sample provides galaxies with the necessary far-IR continuum and \cii~line luminosity measurements; \cii~kinematics; and more importantly, far-IR and \cii-LFs, to undertake the first comprehensive measurement of the gas content of main-sequence galaxies in the post-reionization phase at $z=$~4--6.

To put the molecular gas masses in global context, \citet{ALPINE_YAN20} computed the cosmic molecular gas density using the cosmic-volume-averaged \cii~luminosity density at $z$$\sim$$5$. For the computation of the latter, the derived \cii~LF \citep{ALPINE_YAN20} was integrated down to $7\times10^7\,{\rm L_\odot}$, yielding $\LCII$$\sim$$1.2 \times 10^6\,{\rm L_\odot}$. Several previous studies \citep{ZANELLA18, LIU19} and the study based on the \textit{ALPINE} sample by (see below; \citet{ALPINE_DESSAUGES20}) have derived a linear ratio $\rm M_{molgas}/\LCII~=~\text{15--60}\,M_{\odot}/L_\odot$. Using this ratio, \citet{ALPINE_YAN20} derived the molecular gas mass density $\rho_{\rm molgas}~=~\text{(2--7)}\times10^7\,{\rm M_\odot\,pc^{-3}}$ at $z \sim$~4--6. At a $2\sigma$ and 95\%\ confidence level, the estimated $\rho_{\rm molgas} = \text{(1--14)}\times10^7\,M_\odot\,{\rm Mpc^{-3}}$. The evolution of the cosmic molecular gas mass density (left panel in Figure~\ref{fig:gas_baryons}) closely follows the evolution of the cosmic SFR density (see right panel of Figure~\ref{fig:irx_sfrd}).
Individual molecular gas masses of the \textit{ALPINE} galaxies are calculated in different ways using the abovementioned ingredients. \citet{ALPINE_DESSAUGES20} show promisingly that molecular gas masses derived from the RJ dust continuum, \cii~emission, and dynamical mass agree well with each other within the measurement uncertainties (approx. a factor of two).
The RJ emission at rest-frame $850\,{\upmu \rm{m}}$ was calculated using a far-IR SED template with a temperature of $\sim$$40\,{\rm K}$ normalized to the \textit{ALPINE}/ALMA measurement at rest-frame $150\,{\upmu \rm{m}}$ (see Section~\ref{sec:fir_lf} for more details).
The dynamical mass was calculated for a subset of $18$ nonmerger galaxies classified by \mbox{\citet{ALPINE_LEFEVRE20}} with \cii~emission at S/N$>5$ and with \cii~halo size measurement from \citet{ALPINE_FUJIMOTO20}.
Figure~\ref{fig:gas_baryons} shows the gas fraction and depletion times at $z=$~4--6 derived by \citet{ALPINE_DESSAUGES20} in comparison to measurement from CO emission at lower redshifts from the literature~\citep{SAINTONGE11,BETHERMIN15,LIU19}.
Overall, the rise of the molecular gas fraction traces the evolution of the sSFR (e.g., \citep{FAISST16a,ALPINE_KHUSANOVA20}); it increases with redshift up to $z$$\sim$3--4 (reaching a value of $\sim$$63\%$ at $z$$\sim$$5$) and then flattens out at earlier cosmic times.
Individual measurement of gas masses for \textit{ALPINE} galaxies show a large scatter, which is likely the physical reason for the large range in the measurement of the cosmic molecular gas mass density at $z$$\sim$$5$ (indicated by the vertical size of the boxes in Figure~\ref{fig:gas_baryons}, left panel).
Furthermore, the study finds a continuous decline in the molecular gas depletion time scale ($t_{\rm depl} = M_{\rm gas}/{\rm SFR}$)\endnote{with the total SFR derived from the infrared and UV SFRs.} of a factor of 2--3 from $z=0$ to $z=6$, suggesting a mild enhancement of star-formation efficiency towards high redshifts. This trend is in agreement with the extrapolation of the parametrization by \citet{SCOVILLE17} based on $z<4$ as well as \citet{TACCONI18}, but significantly shallower than $(1+z)^{-1.5}$, which is predicted by semianalytical and cosmological simulations based on the framework of the bathtub model (e.g., \citep{DAVE11,DAVE12,GENEL14,LAGOS15}).  \citet{ALPINE_DESSAUGES20} also show that the depletion time depends on the sSFR (which has been shown to be true also at lower redshifts, e.g.,  \citep{SAINTONGE11}), which may be the reason for the change of depletion time across cosmic time. Still, compared to the upsizing of the gas reservoirs, the increase in star-formation efficiency seems to play a minor role in reaching the high average SFRs at high redshifts, in agreement with the study by \citet{SCOVILLE17}. 

The interpopulation scatter in gas abundances (from $20\%$ to $90\%$) and depletion times (from $10^8\,{\rm yr}$ to $10^9\,{\rm yr}$) may be a result of the interconnection between gas abundance, depletion timescale, and other physical properties, as well as the evolutionary stage of galaxies such as the location on the star-forming main-sequence, metal and dust content, or environment. Such a range of properties is also reflected in the dust attenuation properties (see Section~\ref{sec:sfrd_dust}).

\subsubsection{Enrichment of the IGM at $z$$\sim$4--6}

The comparison of gas content and depletion time to models such as the bathtub model indicates that, not surprisingly, galaxies are not closed boxes but, instead, are governed by complex baryon cycles including inflows and outflows. Constraining this cycle is important for understanding the evolution of galaxies and, closely connected with this, the enrichment of the IGM surrounding them.
An extended dust continuum and \cii~halo around high-redshift galaxies indeed suggests chemical enrichment of the IGM. This was found in multiple $z>4$ galaxies \citep{CARNIANI18,Fujimoto2019} and the \textit{ALPINE} sample allows for the first time a statistical study of extended \cii~halos. As shown by \citet{ALPINE_FUJIMOTO20}, the \cii~halo extent commonly exceeds the UV size of the galaxies by factors of 2--3 or more for galaxies with increasing stellar mass. Furthermore, about $30\%$ of (isolated) galaxies have a \cii~halo component extending more than $10\,{\rm kpc}$, at typical UV sizes of 1--2 kpc \citep{Fujimoto2019}. The trend of increasing \cii~to~UV size ratio with increasing stellar mass and SFR can be explained by star-formation-driven outflows ({\it cf.} simulations by \citet{MAIOLINO15}), although a selection bias cannot be ruled out currently. Deeper observations with ALMA will be needed to confirm this scenario.

On the other hand, \citet{ALPINE_GINOLFI19} suggest direct evidence of star-formation-driven outflows by carefully stacking the \cii~spectra of isolated \textit{ALPINE} galaxies. This study shows an increase in broad wings ($\sim$500--700 km/s) of the \cii~spectra for the most star-forming galaxies in the \textit{ALPINE} sample. This trend is strengthened by a bootstrapping analysis. The calculated mass loading factor (close to unity) is similar to the one of local star-forming galaxies but below the one expected from local AGN (which is approximately $5$ times higher). This suggests that, compared to AGN feedback, star-formation-driven outflows are likely a subdominant mechanism for rapid quenching of normal galaxies in the early Universe. The contribution of AGN to the physics of the most massive \textit{ALPINE} galaxies is still an open question (Barchiesi {et~al.} in prep.).

From these studies, there is strong evidence that star-formation-driven outflows may play an important role in the chemical enrichment of the IGM surrounding the galaxies in the early Universe. In addition, galaxy interactions can contribute to the enrichment, as shown by the triple-merger galaxy system at $z$$\sim$$4.57$ presented in \citet{ALPINE_GINOLFI20}, showing extended \cii~emission covering the IGM between the merging components (see \citet{CAPAK15} for another example at $z$$\sim$$5.5$ of a triple-merger). The IGM enrichment via outflows from stellar feedback and galaxy interactions suggested from these observations is consistent with predictions from recent simulations (see, e.g., \citep{OPPENHEIMER06,OPPENHEIMER10,BOURNAUD11,PALLOTTINI14,NELSON15,ANGLESALCAZAR17,HANI18,GRAZIANI20}); together with galaxy--galaxy interactions, they are largely responsible for the chemical enrichment of the IGM surrounding the galaxies in the early Universe. The triple-merger galaxy system at $z$$\sim$$4.57$ presented in \citet{ALPINE_GINOLFI20} is only one example showcasing this scenario by exhibiting extended \cii~emission covering the IGM between the merging components (see \citet{CAPAK15} for another example at $z$$\sim$$5.5$ of a triple-merger).

\subsection{Contribution of Mergers to Mass Assembly and the Emergence of Rotators} \label{sec:mergers_rotators}

Understanding how galaxies grow at different cosmic times is one of the current outstanding questions. Galaxies can increase their mass via various processes such as internal star formation, the accretion of gas (usually also including minor merging events), and major mergers. It is thought that, specifically at lower redshift, gas accretion and {in situ} star formation are the dominant avenues of mass assembly, with major mergers only playing a subdominant role \citep{Zepf2008,Conselice2014}. On the other hand, several studies suggest mergers to become more important for galaxy evolution at higher redshifts \citep{Kampczyk2013}. This is in line with various simulations suggesting an increase in merger rate towards the early universe~\citep{Gottloeber2001,Rodriguez2015}.

The measurement of the number of major mergers is not trivial. Next to high-resolution imaging, kinematic information (such as about the velocity difference of the merging components) is necessary. Furthermore, approaches focusing on the rest-frame UV or even optical may miss dust-obscured merging components, which leads to an underestimation of the true merger rates. As shown in the previous sections, dust obscuration may be significant even at $z>4$. The triple-merger system examined in \citet{ALPINE_GINOLFI20} is a prime example of such a dust-obscured case that would be missed in UV/optical imaging. The combination of ALMA and sub-kpc resolution HST imaging can mitigate these issues. Importantly, \cii~emission provides kinematic information of UV and optically dust-obscured merger components (see e.g., \citep{ALPINE_JONES21,ALPINE_ROMANO21}).

The \textit{ALPINE} sample allows for the first time a statistical analysis of the merger fraction shortly after the Epoch of Reionization. The galaxies are robustly categorized in mergers and nonmergers (e.g., rotators or dispersion-dominated systems) using the full ALMA \cii~cube in position--velocity space via a morpho-kinematic approach similar to studies of \halpha~at lower redshifts (e.g., \citep{Epinat2008,Law2009}) and by fitting a tilted ring model (see also~\mbox{\citep{Kamphuis2015,JONES17})}. The classification procedures of both methods are detailed in \citet{ALPINE_LEFEVRE20} and \mbox{\citet{ALPINE_JONES21}}. Of the $75$ \cii-detected \textit{ALPINE} galaxies, \citet{ALPINE_ROMANO21} identified $23$ of them as major merger systems in the mass range of $\logm$ = 9--11 with a mass ratio less than $4$, a velocity difference between the components of less than $500\,{\rm km/s}$, and a physical distance of less than $20\,{\rm kpc}$. The authors computed the fraction of major mergers in this sample ($\sim$$45\%$ and $\sim$$35\%$ for $z$$\sim$$4.5$ and $z$$\sim$$5.5$, respectively). These values are a factor of two higher than at $z$ = 2--3 ($\sim$$20\%$) and other studies at the same redshift using only HST imaging (e.g., \citep{CONSELICE09}). The latter could highlight an incompleteness due to missing dust-obscured merger components in the UV and optical data. The computed fraction of major mergers is consistent with that predicted by the \textit{Evolution and Assembly of Galaxies and their Environments} (EAGLE) hydrodynamical simulation \citep{SCHAYE15}.
The major merger fraction of 35--45\% derived from the \textit{ALPINE} sample by \citet{ALPINE_ROMANO21} suggests that major mergers could play a significant role in galaxy evolution at these redshifts. To estimate the contribution to the galaxy mass assembly, a merger timescale has to be assumed, which comes with significant uncertainty. Assuming typical merger timescale representations by \citet{KITZBICHLER08}, \citet{JIANG14}, and \citet{SNYDER17}, the contribution of major mergers to the SFR density at $z=$~4--6 varies from less than $5\%$ (for Kitzbichler et~al.) to nearly $30\%$ (for Snyder et~al.).

The recent studies by \citet{ALPINE_LEFEVRE20} and \citet{ALPINE_JONES21} picture the morphological diversity of the \textit{ALPINE} sample at $z=$~4--6. Specifically, next to the approximately $40\%$ interacting galaxies, about $30\%$ are classified as dispersion-dominated systems and about $\sim$$15\%$ are found to be rotators. The rotators are of special interest as they provide constraints on the dark matter mass of galaxies and also suggest recent relatively smooth mass growth in a cold-gas-accretion-dominated scenario.
Rotators have been characterized at \mbox{$z<4$} mainly through Integral Field Unit (IFU) observations of the \halpha~emission line (\mbox{e.g.,~\citep{EPINAT08,FORSTERSCHREIBER09,GNERUCCI11}}), but also via \cii~emission for a handful of bright starburst and quasars at higher redshifts (e.g., \citep{JONES17,PENSABENE20,TADAKI20,FRATERNALI21,LELLI21,RIZZO20,RIZZO21}). Next to a main-sequence galaxy at $z=4.2$ \citep{NEELEMAN20} and two galaxies at $z>6$ \citep{SMIT18,BAKX20}, the \textit{ALPINE sample} complements these studies with typical main-sequence galaxies at $z=$~4--6.
Within the \textit{ALPINE} sample, \citet{ALPINE_JONES21} find rotators with rotational velocities between 50--250 km/s and extended rotational structures commonly less than $7\,{\rm kpc}$ (although some out to $\sim$$10\,{\rm kpc}$) (see also \citep{ALPINE_FUJIMOTO20}). The flattening of the rotation curves at a radius of $\sim$1--2 kpc suggest dominating baryon fractions over dark matter. However, strong conclusions cannot be drawn because of insufficient spatial resolution of the ALMA observations. A spatial resolution of at least $\sim$$1\,{\rm kpc}$ (or 0.1--0.2$\arcsec$) at a high sensitivity out to large radii ($\sim$7--10  kpc) is necessary for definite answers (see Figure 7 in \citep{ALPINE_JONES21}).

\section{Discussion}\label{sec:discussion}

\textit{ALPINE} is a survey that is well-placed to study the ramp-up of galaxy evolution at redshifts of $z=$~4--6 and has led to many important science results, as described in the previous sections. While past studies have focused on bright sub-mm galaxies or quasars, mostly due to sensitivity limitations, \textit{ALPINE} has broadened the view to more typical main-sequence star-forming galaxies in the post-reionization era. 

The large sample size allowed the first statistical comparison of \cii~emission density of a post-reionization sample and the local Universe \citep{ALPINE_YAN20}. Compared to the currently best local comparison sample \citep{HEMMATI17}, galaxies at $z=$~4--6 show statistically an excess of \cii~emission at the luminous end of the \cii~luminosity function \citep{ALPINE_YAN20}. Since \cii~is a common cooling line in star-forming galaxies, the excess of luminous \cii~emitters in high-redshift main-sequence galaxies is not surprising due to their overall enhanced rates of (specific) star-formation compared to main-sequence galaxies at lower redshifts. \citet{ALPINE_LOIACONO21} found that clustering of \cii~emitters adds a significant uncertainty to the measured \cii~emission density. Specifically, it is intriguing that $7$ out of $8$ unambiguously identified serendipitous $z>4$ \cii~emitters are within less than $750\,{\rm km/s}$ ($\Delta z < 0.015$) in velocity space and <$17\arcsec$ in projected radius\endnote{Radius from phase center where primary beam attenuation is <$90\%$.} of the main target \citep{ALPINE_LOIACONO21}. This suggests significant spatial clustering of \cii~emitters in the early Universe and may be used to inform cosmological structure formation models in the future.

The excess of strong \cii~emitters at high redshifts may be related to their increased star formation. This is mainly motivated by the strong correlation between \cii~and star formation found in galaxy samples up to $z$$\sim$$2$ (e.g., \citep{PINEDA13,HERRERACAMUS15,DELOOZE11,DELOOZE14}). Although verified at lower redshifts, it was far from clear how well \cii~traces star formation at early cosmic epochs. Several studies suggested a \cii~deficit in high-$z$ galaxies (e.g., \citep{MAIOLINO15,WILLOTT15,PENTERICCI16,BRADAC17,CARNIANI18}), resulting in a large scatter in the \cii$-$SFR relation. Furthermore, the different places of origin of \cii~(from \hii~regions, the diffuse neutral and ionized ISM, or photo-dissociation regions) could further be relevant for its connection to physical properties of the galaxies. The \textit{ALPINE} sample emphasized the strong relation between \cii~and total SFR ({i.e.}, obscured and unobscured) and also verified its similarity to local galaxies \citep{ALPINE_SCHAERER20}.
However, there are still open questions. As mentioned several times in this article, the \textit{ALPINE} galaxies reside on the more massive end of the star-forming main-sequence and are also suggested to be already substantially metal-enriched \citep{ALPINE_FAISST20}. Their ISM properties are therefore not substantially different from their descendants at lower redshifts (see also \citep{ALPINE_VANDERHOOF22}). A way to remedy this is to build up a statistical sample of low-mass (likely lensed) galaxies at $z$$\sim$4--6 to study \cii~physics in truly low-metallicity environments. Furthermore, carrying out follow-up observations with ALMA of other far-IR lines (such as \nii~at $122\,\upmu{\rm m}$ or $205\,\upmu{\rm m}$) would allow to statistically pinpoint the dominant origin of \cii~emission in these galaxies (see observations and analyses by  \citep{PAVESI16,PAVESI19}). Cospatial resolution-matched observations of optical emission (such as \halpha) from JWST and \cii~from ALMA would further help to investigate this issue.

As the \textit{ALPINE} galaxies are rather mature in terms of their stellar mass and ISM properties, it might not be very surprising to find a relatively large dust attenuation in this population. The most massive galaxies in the sample have a similar fraction of dust-obscured star formation as galaxies at the peak of cosmic SFR density \citep{ALPINE_FUDAMOTO20}. The measurement of dust mass at this early epoch places constraints on the dust production mechanisms and chemical evolution models. The analysis of ages and dust masses by \citet{ALPINE_POZZI21} draws the picture that these galaxies are likely the progenitors of bulge-dominated spiral or elliptical galaxies at lower redshifts. The dust production has to occur quickly, in line with the evolution of proto-spheroids \citep{CALURA17}. Dust production by supernovae on short time scales provide the most likely scenario for the high dust masses of these $z=$~4--6 galaxies \citep{Gall2018,ALPINE_BURGARELLA22}. However, it is interesting that current models have difficulties to reproduce the high dust-mass to stellar-mass ratios ($\geq$$0.03$) of young $z$$\sim$5--6 galaxies with high sSFRs ($\geq 40\,{\rm Gyr^{-1}}$) \citep{ALPINE_BURGARELLA22}. This becomes even more intriguing with the discovery of entirely optically dark galaxies at $z$$\sim$$6.7$ and $z$$\sim$$7.4$ \citep{FUDAMOTO21,INAMI22} from the \textit{REBELS} sample (the $z>6$ continuation of \textit{ALPINE}, see survey overview described in \citet{BOUWENS21}), although there could be significant uncertainties in the stellar masses (hence, the dust to stellar mass ratios) of these galaxies (Sommogivo {et~al.} in prep.). Such dust-obscured galaxies at high redshifts challenge current dust formation models and at the same time suggest that rest-UV observations are still missing a substantial amount ($\sim$10--25\%) of galaxies contributing to the cosmic SFR density at $z>6$ \citep{FUDAMOTO21}.

Several studies have consistently found a continuous increase in sSFR of galaxies out to $z$$\sim$$3$, based on robust, optical, and infrared-based total SFR measurements \citep{ILBERT15}. This trend was found to continue at $z>3$, although there was no consensus on the steepness of the sSFR increase. While some studies suggested a flattening \citep{BOUWENS12,TASCA15,DAVIDZON18}, others found a more significant increase \citep{FAISST16a,STARK13,DEBARROS14}, which could be due to the use of different SFR tracers. The study by \citet{ALPINE_KHUSANOVA20} provided the first robust measurement of star formation at $z$$\sim$$5$ from ALMA and supports a flattening of the sSFR evolution. 
The apparent inconsistency in slope is likely due to the use of different star-formation indicators. For example, the study by \citet{FAISST16a} used \halpha~(derived from Spitzer colors) as tracer for star formation (hence, sSFR) on short timescales ($\sim$$10\,{\rm Myrs}$) and found a steeper slope ({i.e.}, higher sSFR at a given redshift) than other studies using continuum-based, $100\,{\rm Myr}$-averaged SFR measurements (e.g., from SED fitting or far-IR SFR as in \citep{ALPINE_KHUSANOVA20}). Short-term SFR measurement may be significantly increased due to burstiness in the star-formation process at high redshifts as suggested in \citet{FAISST19b}, and thus, could explain the higher~sSFRs.

A higher gas fraction or higher star-formation efficiency could both play a role in increasing the star-formation activity of $z>4$ galaxies. The results from \citet{ALPINE_DESSAUGES20} on $z>4$ galaxies show that, compared to low-$z$ galaxies, on average the gas fractions are increased by a factor of $6$ ($10\%$ vs. $60\%$), while the depletion times (i.e., inverse star-formation efficiency) are shorter by a factor of two ($1\,{\rm Gyr}$ vs. $0.5\,{\rm Gyrs}$). The gas fraction traces closely the sSFR evolution in shape (including flattening at $z>3$). The decrease in depletion time is significantly less than that predicted by the ``bathtub'' model, in which main-sequence galaxies lie in a quasi-equilibrium state \citep{BOUCHE10,DAVE12,LILLY13,DEKEL14}, but is consistent with extrapolated CO measurements from \citep{TACCONI18} and the relative change in \mbox{\citet{SCOVILLE17}} derived from RJ dust-continuum gas masses. This suggests that $z$$\sim$$5$ galaxies (on the main-sequence) are not significantly more efficient in forming stars than present-day galaxies (a conclusion also obtained by \citet{SCOVILLE17}). Most of the redshift evolution of star formation on the main sequence is therefore due to increased gas reservoirs rather than higher star-formation efficiency. The latter, however, may be more important in pushing galaxies at a given stellar mass and redshift above the main-sequence, and so creating the observed starburst population \citep{SCOVILLE17,SILVERMAN15}.
The high gas fraction and the slowly changing depletion times suggest that a significant amount of gas is expelled from galaxies via outflows. As shown by \citep{ALPINE_GINOLFI19}, outflows are ubiquitous in highly star-forming $z>4$ galaxies and may also be responsible for extended \cii~halos \citep{ALPINE_FUJIMOTO20}.

The contribution of mergers to the mass growth of galaxies at $z>4$ is still not clear. Although \textit{ALPINE} provides currently the best constraints of the merger fraction thanks to dust-independent measurements from far-IR continuum and \cii~emission, the translation to a merger rate and mass assembly due to major mergers depends strongly on the assumption of merger timescales. Assuming current state-of-the-art estimates of merger timescales, the contribution of major mergers to galaxy mass growth can be up to $30\%$. Clearly, there is more work to be done to understand the merging process---either through more sophisticated simulation work or enhanced statistics on the merger stages (early, late mergers) in the population.

\section{Conclusions}\label{sec:end}

\textit{ALPINE} has progressed our understanding of galaxy evolution by measuring the dust and gas properties of galaxies at $z=$~4--6 through sub-mm observations. 
Most previous observations were based on luminous sub-mm galaxies or only a handful of main-sequence galaxies at these redshifts. Furthermore, rest-frame UV-based surveys only provide a limited view on the total galaxy population as they miss significant amounts of dust-obscured galaxies.
Combined with UV to optical imaging and spectroscopic data, \textit{ALPINE} is currently the largest multiwavelength sample of main-sequence galaxies in the post-reionization era and builds a basis for future observations.
The new findings enabled by \textit{ALPINE} can be summarized as follows:

\begin{itemize}

\item The demographics of $z$$\sim$$5$ \cii-emitters, {i.e.}, LF, is quantified for the first time by the \textit{ALPINE} main and serendipitous samples. This study shows evidence of a possible much higher volume density of luminous \cii-emitters at $z$$\sim$$5$ than that of $z$$\sim$$0$.
\item The \cii~emission is a good tracer of total SFR for $\logm > 9$ main-sequence galaxies at $z$$\sim$$5$. The relation is consistent with what was found at lower redshifts (including local galaxies). This suggests no significant evolution of the \cii$-$SFR relation across cosmic time up to $z=6$ for main-sequence galaxies at this stellar mass.
\item Significant dust attenuation (45--65\%) is found in the most massive ($\logm>10$) galaxies, together with a significant number of ``UV/optical-dark'' galaxies at $z=$~4--6. This shows that UV-selected samples miss a dusty population of galaxies even at $z>4$. The most massive galaxies still challenge current models of dust production.
\item The gas fraction of main-sequence galaxies increases out to $z=6$ and follows in shape the evolution of the sSFR. While the gas fraction evolves significantly, the star-formation efficiency does not; hence, it is suggested to contribute little to the increased star-formation activity in high-$z$ galaxies.
\item Outflows are ubiquitous in highly star-forming main-sequence galaxies and may contribute to enriching the IGM with metals. Enriched IGM is also suggested by extended \cii~emission halos around the galaxies.
\item The morpho-kinematic selection of major mergers suggests an increase of galaxy interactions at $z=$~4--6. Adopting state-of-the-art estimates of merger timescales, major mergers contribute at most $30\%$ to the galaxy mass assembly at that epoch.
\end{itemize}

Despite these advances, future research is important to understand better the details of this galaxy population and to connect it with galaxies in the Epoch of Reionization.
While the above results contribute to our understanding of relatively massive and mature main-sequence galaxies, conclusions about the fainter end of the population cannot be made with the current sample. Extending the sample to lower stellar masses and (likely) less mature galaxies (in terms of chemical enrichment) would provide an important comparison. For example, does the \cii$-$SFR relations still hold for such galaxies? Is star formation more dominated by cold gas accretion or mergers in fainter galaxies?
A large sample of lensed galaxies at $z=$~4--6 would serve as a promising avenue to answer such questions. With enough magnification, ALMA observations of the low-mass end of the main-sequence would not require significantly more time.
\textit{ALPINE} builds a basis sample at this epoch, which can serve as ``anchor'' for other studies at lower and higher redshifts. An important study to mention in this context is \textit{REBELS}, a high-redshift extension of \textit{ALPINE} to measure \cii~and far-IR continuum at $z>6.5$ using a spectral scan of galaxies with photometric redshifts \citep{BOUWENS21}. A careful comparison between \textit{REBELS} and \textit{ALPINE} would inform us about possible different galaxy evolution in the Epoch of Reionization as well as shortly thereafter (see, e.g., \citep{SOMMOVIGO22}). 
In addition to existing surveys, a higher-resolution \cii~and dust-continuum follow-up with ALMA would further help us understand this important epoch in galaxy evolution. Sub-kpc resolution observations would reveal the location of star-formation as well as their gas velocity dispersion. The comparison to studies at lower redshifts \citep{DESSAUGES19} would help us constrain star formation in general as well as the role of feedback from stellar winds, SNe, and AGN on smaller spatial scales than currently possible. 
The combination of deep high-resolution and shallow low-resolution observations with ALMA would also significantly improve the measurement of velocity curves to pin down the dark matter content of these galaxies. The spatial sampling of current observations is not fine enough to unambiguously determine between a flat or declining velocity curve out to large radii \citep{ALPINE_JONES21}.
Finally, JWST will be the workhorse for optical exploration of the galaxies, by providing, for example, resolved metallicity measurements or stellar mass and optical dust maps. The paper by \citet{ALPINE_VANDERHOOF22} already provides a taste of the possibilities that arise by combining optical emission lines with UV and sub-mm measurements.

\vspace{6pt} 
\authorcontributions{Project administration, Olivier LeF\`evre; Writing -- original draft preparation, Andreas Faisst and Lin Yan; Writing -- review and editing, Matthieu B\'ethermin, Paolo Cassata, Miroslava Dessauges-Zavadsky, Yoshinobu Fudamoto, Michele Ginolfi, Carlotta Gruppioni, Gareth Jones, Yana Khusanova, Francesca Pozzi, Michael Romano, Daniel Schaerer, John Silverman and Brittany Vanderhoof. All authors have read and agreed to the published version of the manuscript.}

\funding{
This research received no external funding.
}
\institutionalreview{Not applicable.}

\informedconsent{Not applicable.}

\dataavailability{
The data used in this article are available online at \url{https://cesam.lam.fr/a2c2s/data_release.php}, accessed on May 31, 2022
.
} 

\acknowledgments{
We thank the reviewers for their useful comments, which greatly improved this article.
 \textls[-15]{M.R. acknowledges support from the Narodowe Centrum Nauki (UMO-2020/38/E/ST9/00077).} G.C.J. acknowledges ERC Advanced Grant 695671 “QUENCH” and support by the Science and Technology Facilities Council (STFC). YK acknowledges the support by funding from the European Research Council Advanced Grant ERC–2010–AdG–268107–EARLY.
This paper makes use of the following ALMA data: ADS/JAO.ALMA\#2017.1.00428.L. ALMA is a partnership of ESO (representing its member states), NSF (USA) and NINS (Japan), together with NRC (Canada), MOST and ASIAA (Taiwan), and KASI (Republic of Korea), in cooperation with the Republic of Chile. The Joint ALMA Observatory is operated by ESO, AUI/NRAO and NAOJ.
This research made use of Astropy (\url{http://www.astropy.org}, accessed on May 31, 2022), a community-developed core Python package for Astronomy~\citep{ASTROPY13,ASTROPY18}.
We would also like to recognize the contributions from all of the members of the COSMOS Team who helped in obtaining and reducing the large amount of multiwavelength data that are now publicly available through IRSA at \url{http://irsa.ipac.caltech.edu/Missions/cosmos.
html}, accessed on May 31, 2022.
This article is dedicated to the memory of Olivier Le F\`evre, PI of the ALPINE survey.
}

\conflictsofinterest{
The authors declare no conflict of interest.
}

\begin{adjustwidth}{-\extralength}{0cm}
\printendnotes[custom]
\reftitle{References}

\end{adjustwidth}

\end{document}